\RequirePackage{lineno}

\documentclass[aps,prd,twocolumn,superscriptaddress,showpacs]{revtex4}

\usepackage{graphicx}  
\usepackage{dcolumn}   
\usepackage{bm}        
\usepackage{amssymb}   

\usepackage{xspace}

\def\etal{{\sl et al.}}
\def\etjj{{e\tau jj}}
\def\mtjj{{\mu\tau jj}}
\def\ltjj{{\ell\tau jj}}
\def\mt0{{\mu\tau 01}}
\def\gsim{\mathrel{\rlap{\raise.4ex\hbox{$>$}} {\lower.6ex\hbox{$\sim$}}}} 
\def\lsim{\mathrel{\rlap{\raise.4ex\hbox{$<$}} {\lower.6ex\hbox{$\sim$}}}}

\newcommand{\tautau}{\mbox{\ensuremath{\tau\tau}}}

\newcommand{\met}{\mbox{\ensuremath{\slash\kern-.7emE_{T}}}}
\newcommand{\mht}{\mbox{\ensuremath{\slash\kern-.7emH_{T}}}}
\newcommand{\mpt}{\mbox{\ensuremath{\slash\kern-.5emT_{T}}}}

\newcommand{\ttbar}{\mbox{\ensuremath{t\overline t}}}
\newcommand{\wj}{\mbox{\ensuremath{W+\mathrm{jets}}}}
\newcommand{\zj}{\mbox{\ensuremath{Z+\mathrm{jets}}}}

\begin{document}


\begin{flushleft}

{FERMILAB-PUB-12-621-E} \\

\end{flushleft}

\title{Search for the Higgs boson in lepton, tau and jets final states}


%
\affiliation{LAFEX, Centro Brasileiro de Pesquisas F\'{i}sicas, Rio de Janeiro, Brazil}
\affiliation{Universidade do Estado do Rio de Janeiro, Rio de Janeiro, Brazil}
\affiliation{Universidade Federal do ABC, Santo Andr\'e, Brazil}
\affiliation{University of Science and Technology of China, Hefei, People's Republic of China}
\affiliation{Universidad de los Andes, Bogot\'a, Colombia}
\affiliation{Charles University, Faculty of Mathematics and Physics, Center for Particle Physics, Prague, Czech Republic}
\affiliation{Czech Technical University in Prague, Prague, Czech Republic}
\affiliation{Center for Particle Physics, Institute of Physics, Academy of Sciences of the Czech Republic, Prague, Czech Republic}
\affiliation{Universidad San Francisco de Quito, Quito, Ecuador}
\affiliation{LPC, Universit\'e Blaise Pascal, CNRS/IN2P3, Clermont, France}
\affiliation{LPSC, Universit\'e Joseph Fourier Grenoble 1, CNRS/IN2P3, Institut National Polytechnique de Grenoble, Grenoble, France}
\affiliation{CPPM, Aix-Marseille Universit\'e, CNRS/IN2P3, Marseille, France}
\affiliation{LAL, Universit\'e Paris-Sud, CNRS/IN2P3, Orsay, France}
\affiliation{LPNHE, Universit\'es Paris VI and VII, CNRS/IN2P3, Paris, France}
\affiliation{CEA, Irfu, SPP, Saclay, France}
\affiliation{IPHC, Universit\'e de Strasbourg, CNRS/IN2P3, Strasbourg, France}
\affiliation{IPNL, Universit\'e Lyon 1, CNRS/IN2P3, Villeurbanne, France and Universit\'e de Lyon, Lyon, France}
\affiliation{III. Physikalisches Institut A, RWTH Aachen University, Aachen, Germany}
\affiliation{Physikalisches Institut, Universit\"at Freiburg, Freiburg, Germany}
\affiliation{II. Physikalisches Institut, Georg-August-Universit\"at G\"ottingen, G\"ottingen, Germany}
\affiliation{Institut f\"ur Physik, Universit\"at Mainz, Mainz, Germany}
\affiliation{Ludwig-Maximilians-Universit\"at M\"unchen, M\"unchen, Germany}
\affiliation{Fachbereich Physik, Bergische Universit\"at Wuppertal, Wuppertal, Germany}
\affiliation{Panjab University, Chandigarh, India}
\affiliation{Delhi University, Delhi, India}
\affiliation{Tata Institute of Fundamental Research, Mumbai, India}
\affiliation{University College Dublin, Dublin, Ireland}
\affiliation{Korea Detector Laboratory, Korea University, Seoul, Korea}
\affiliation{CINVESTAV, Mexico City, Mexico}
\affiliation{Nikhef, Science Park, Amsterdam, the Netherlands}
\affiliation{Radboud University Nijmegen, Nijmegen, the Netherlands}
\affiliation{Joint Institute for Nuclear Research, Dubna, Russia}
\affiliation{Institute for Theoretical and Experimental Physics, Moscow, Russia}
\affiliation{Moscow State University, Moscow, Russia}
\affiliation{Institute for High Energy Physics, Protvino, Russia}
\affiliation{Petersburg Nuclear Physics Institute, St. Petersburg, Russia}
\affiliation{Instituci\'{o} Catalana de Recerca i Estudis Avan\c{c}ats (ICREA) and Institut de F\'{i}sica d'Altes Energies (IFAE), Barcelona, Spain}
\affiliation{Uppsala University, Uppsala, Sweden}
\affiliation{Lancaster University, Lancaster LA1 4YB, United Kingdom}
\affiliation{Imperial College London, London SW7 2AZ, United Kingdom}
\affiliation{The University of Manchester, Manchester M13 9PL, United Kingdom}
\affiliation{University of Arizona, Tucson, Arizona 85721, USA}
\affiliation{University of California Riverside, Riverside, California 92521, USA}
\affiliation{Florida State University, Tallahassee, Florida 32306, USA}
\affiliation{Fermi National Accelerator Laboratory, Batavia, Illinois 60510, USA}
\affiliation{University of Illinois at Chicago, Chicago, Illinois 60607, USA}
\affiliation{Northern Illinois University, DeKalb, Illinois 60115, USA}
\affiliation{Northwestern University, Evanston, Illinois 60208, USA}
\affiliation{Indiana University, Bloomington, Indiana 47405, USA}
\affiliation{Purdue University Calumet, Hammond, Indiana 46323, USA}
\affiliation{University of Notre Dame, Notre Dame, Indiana 46556, USA}
\affiliation{Iowa State University, Ames, Iowa 50011, USA}
\affiliation{University of Kansas, Lawrence, Kansas 66045, USA}
\affiliation{Kansas State University, Manhattan, Kansas 66506, USA}
\affiliation{Louisiana Tech University, Ruston, Louisiana 71272, USA}
\affiliation{Northeastern University, Boston, Massachusetts 02115, USA}
\affiliation{University of Michigan, Ann Arbor, Michigan 48109, USA}
\affiliation{Michigan State University, East Lansing, Michigan 48824, USA}
\affiliation{University of Mississippi, University, Mississippi 38677, USA}
\affiliation{University of Nebraska, Lincoln, Nebraska 68588, USA}
\affiliation{Rutgers University, Piscataway, New Jersey 08855, USA}
\affiliation{Princeton University, Princeton, New Jersey 08544, USA}
\affiliation{State University of New York, Buffalo, New York 14260, USA}
\affiliation{University of Rochester, Rochester, New York 14627, USA}
\affiliation{State University of New York, Stony Brook, New York 11794, USA}
\affiliation{Brookhaven National Laboratory, Upton, New York 11973, USA}
\affiliation{Langston University, Langston, Oklahoma 73050, USA}
\affiliation{University of Oklahoma, Norman, Oklahoma 73019, USA}
\affiliation{Oklahoma State University, Stillwater, Oklahoma 74078, USA}
\affiliation{Brown University, Providence, Rhode Island 02912, USA}
\affiliation{University of Texas, Arlington, Texas 76019, USA}
\affiliation{Southern Methodist University, Dallas, Texas 75275, USA}
\affiliation{Rice University, Houston, Texas 77005, USA}
\affiliation{University of Virginia, Charlottesville, Virginia 22904, USA}
\affiliation{University of Washington, Seattle, Washington 98195, USA}
\author{V.M.~Abazov} \affiliation{Joint Institute for Nuclear Research, Dubna, Russia}
\author{B.~Abbott} \affiliation{University of Oklahoma, Norman, Oklahoma 73019, USA}
\author{B.S.~Acharya} \affiliation{Tata Institute of Fundamental Research, Mumbai, India}
\author{M.~Adams} \affiliation{University of Illinois at Chicago, Chicago, Illinois 60607, USA}
\author{T.~Adams} \affiliation{Florida State University, Tallahassee, Florida 32306, USA}
\author{G.D.~Alexeev} \affiliation{Joint Institute for Nuclear Research, Dubna, Russia}
\author{G.~Alkhazov} \affiliation{Petersburg Nuclear Physics Institute, St. Petersburg, Russia}
\author{A.~Alton$^{a}$} \affiliation{University of Michigan, Ann Arbor, Michigan 48109, USA}
\author{A.~Askew} \affiliation{Florida State University, Tallahassee, Florida 32306, USA}
\author{S.~Atkins} \affiliation{Louisiana Tech University, Ruston, Louisiana 71272, USA}
% ...
%
%
\collaboration{The D0 Collaboration\footnote{with visitors from
$^{a}$Augustana College, Sioux Falls, SD, USA,
$^{b}$The University of Liverpool, Liverpool, UK,
$^{c}$UPIITA-IPN, Mexico City, Mexico,
$^{d}$DESY, Hamburg, Germany,
$^{e}$SLAC, Menlo Park, CA, USA,
$^{f}$University College London, London, UK,
$^{g}$Centro de Investigacion en Computacion - IPN, Mexico City, Mexico,
$^{h}$ECFM, Universidad Autonoma de Sinaloa, Culiac\'an, Mexico
and
$^{i}$Universidade Estadual Paulista, S\~ao Paulo, Brazil.
}} \noaffiliation
\vskip 0.25cm

\date{\today}

\vspace*{2.0cm}

\begin{abstract}
We present a search for the standard model Higgs boson in final states
with an electron or muon and a hadronically decaying tau lepton in association with 
two or more jets using 9.7 fb$^{-1}$ of Run II Fermilab Tevatron Collider
data collected with the D0 detector.  The analysis is sensitive to
Higgs boson production via gluon fusion, associated vector boson 
production, and vector boson fusion, followed by the Higgs boson decay to tau lepton pairs 
or to $W$ boson 
pairs.  The ratios of 95\% C.L. upper limits on the 
cross section times branching ratio to those predicted by the standard model are obtained for
orthogonal subsamples that are enriched in either $H\rightarrow \tautau$ decays or $H\rightarrow WW$ 
decays, and 
for the combination of these subsample limits.  
The observed and expected limit ratios for 
the combined subsamples at a Higgs boson mass of 125 GeV are 11.3 and 9.0 respectively.
\end{abstract}


\maketitle


\newpage

\section{\label{sec-intro}Introduction}

The standard model (SM) of particle physics postulates a complex Higgs doublet field as the source of
electroweak symmetry breaking, giving rise to non-zero masses of the 
$W$ and $Z$ vector bosons and the 
fundamental fermions.  The mass of the spin-zero Higgs boson, $H$, that survives after the symmetry
breaking is not predicted by the SM, but is constrained 
by direct searches at the LEP~\cite{lep-higgs}, Tevatron~~\cite{tev-ichep-combo} and
LHC~\cite{atlas-higgs,cms-higgs} colliders, and 
by precision electroweak measurements~\cite{gfitter} to be in the range 122 -- 127 GeV
at the 95\% C.L.  
The boson observed at a mass of about 126 GeV by ATLAS and CMS~\cite{atlas-higgs,cms-higgs}
when combining evidence for a narrow resonance in the $\gamma\gamma$ and
$ZZ$ channels has production and decay properties that are
consistent with the SM Higgs boson, given the current sensitivities.
Evidence for a particle compatible with the discovered boson with the decay $b\overline b$
has been reported independently by the Tevatron 
experiments~\cite{d0-combo, cdf-combo, tev-combo}.
It is now important to measure its couplings for all accessible 
particles, in particular to leptons for which no evidence currently exists.

In this paper
we present a search for the Higgs boson in final states
that are sensitive to both the $H\rightarrow \tau\tau$ and $H\rightarrow WW$ decay modes
containing a hadronically decaying tau lepton, a muon or electron plus at least two jets.
We use data collected with the D0 detector at the Fermilab proton-antiproton 
Tevatron Collider at $\sqrt s=1.96$ TeV.
The search is conducted at ten Higgs boson masses ($M_H$) between 105 and 150 GeV
in 5 GeV intervals.
In the following $\tau_h$ represents a hadronically decaying tau lepton, 
$\ell$ denotes a lepton ($\mu$ or $e$) and $j$ represents hadronic jets.
The symbols ``$\mtjj$'' and ``$\etjj$'' denote the two individual search channels,
collectively described as ``$\ltjj$''.

Over the mass range $115 \leq M_H \leq 150$ GeV
the Higgs boson decay branching fractions 
vary considerably, with $H\rightarrow b\overline b$  
being the dominant decay for $M_H \lsim 135$ GeV
and $H\rightarrow W^+W^-$ becoming important for $M_H \gsim 135$ GeV.  
Below $M_H \lsim 125$ GeV, $H\rightarrow \tau^+ \tau^-$ has
an appreciable ($\approx 8\%$) branching fraction.
Previous analyses by the D0 and CDF Collaborations have mainly 
focused on the decay modes $H\rightarrow b\overline b$ in the 
low mass region~\cite{d0-combo, cdf-combo, tev-combo}
and 
$H\rightarrow WW$ with both $W$ bosons decaying to a lepton and neutrino in the high mass
region~\cite{tev-ichep-combo}.  
A  D0 publication~\cite{ttjj_old} reported a Higgs boson search in 
the $\tau_h \mu$ final state with zero or one jet, and the 
$\mtjj$  and $\etjj$ final states,
using 7.3, 6.2 and 4.3 fb$^{-1}$ of data respectively.  
The CDF collaboration has published a search in the $\tau_h \ell ~+ \geq 1$ jet 
final state using 6.0 fb$^{-1}$
of data~\cite{cdf_tautau}.
Here we report updated results for both the $\mtjj$ and $\etjj$ 
searches with the full D0 9.7 fb$^{-1}$ 2002 -- 2011
data set with an improved analysis using an extended set of variables for discriminating
signal and background and an improved multivariate analysis.  The current $\ltjj$ analyses 
supersede those of Ref.~\cite{ttjj_old}.

The Higgs boson production processes considered are 
({\it i})  gluon fusion (GF), $gg \rightarrow H$ (+ jets);
({\it ii}) vector boson fusion (VBF), $q \overline q \rightarrow q\overline q H$;
({\it iii}) associated vector boson and Higgs boson production (VH),
$q \overline q \rightarrow VH$, where $V$ is a $W$ or $Z$ boson, 
and $V\rightarrow q\overline q$; and
({\it iv}) associated Higgs boson and $Z$ boson production (HZ),
$q \overline q \rightarrow HZ$, with $H\rightarrow b\overline b$ and $Z\rightarrow \tau\tau$.
The GF, VBF, and VH processes are further subdivided according to the Higgs boson
decay, $H\rightarrow \tau\tau$ or $H\rightarrow WW$,
and these subchannels are denoted as GF$_{\tau\tau}$, GF$_{WW}$, etc.

Tau leptons can occur at lower $M_H$ through direct decays of the Higgs boson 
or, at higher $M_H$, indirectly 
from $H\rightarrow VV $ with $V\rightarrow \tau +X$.
The leptons may arise from $\tau$ decay or, at high $M_H$, directly from $V$ decay.
Thus the $\ltjj$ channels 
are more uniformly sensitive to Higgs boson
production over the full mass range investigated than are the dedicated
$H\rightarrow b\overline b$ or 
$H\rightarrow WW\rightarrow \ell\overline \ell \nu \overline \nu$ analyses, 
thus improving the combined search sensitivity 
particularly in the intermediate mass region around 135 GeV.

\section{\label{sec-detector}The D0 detector }

The D0 detector \cite{nim1, nim2, nimmu} contains tracking, calorimeter and muon  
subdetector systems. 
Silicon microstrip tracking detectors (SMT) near the interaction point cover 
pseudorapidity $| \eta | < 3$ to provide tracking and vertexing information.
The SMT~\cite{smt,smtlzero} 
contains cylindrical barrel layers aligned with their axes parallel to the beams 
and disk segments.
The disks are perpendicular to the beam
axis, interleaved with, and extending beyond, the barrels.  
The central fiber tracker (CFT) surrounds the SMT, providing coverage to about
$| \eta | = 2$.
The CFT has eight concentric cylindrical layers of overlapped scintillating fibers
providing axial and stereo ($\pm 3^\circ$) measurements. 
A 1.9 T solenoid surrounds these tracking detectors.

Three uranium liquid-argon calorimeters measure particle energies. 
The central calorimeter (CC) covers $| \eta | < 1$, and two end calorimeters (EC) extend 
coverage to about $| \eta | = 4$. 
The calorimeter is highly segmented along the particle direction, 
with four electromagnetic (EM) and four or five hadronic sections in depth, 
and segmentation transverse to the particle direction 
with typically $\Delta\eta = \Delta\phi = 0.1$, 
where $\phi $ is the azimuthal angle ($\Delta\eta = \Delta\phi = 0.05$ in the third EM depth segment).
The calorimeters are supplemented with central and forward scintillating strip 
preshower detectors (CPS and FPS) located in front of the CC and EC.
Intercryostat detectors (ICD) provide added sampling in the region
$1.1 < |\eta | < 1.4$ where the CC and EC cryostat walls degrade the 
calorimeter energy resolution.

Muons are measured just outside the calorimeters, and twice more
outside 1.8 T iron toroidal magnets, over the range $| \eta | < 2$. 
Each measurement is based on scintillation counters and several layers of tracking chambers.
Scintillators surrounding the exiting beams allow determination of the 
luminosity~\cite{lumi}.

A three level trigger system selects events for data logging at about 100 Hz.
The first level trigger (L1) is based on fast custom logic for several subdetectors and 
is capable of making decisions after each beam crossing.  
The second level trigger (L2) makes microprocessor-based decisions using multi-detector information.  
The third level trigger (L3) uses fully digitized
outputs from all detectors to refine the decision and select events for offline processing.

The data collected for this analysis comes from the full Tevatron Run II period extending
from 2002 to 2011.  The data set is divided into two epochs, Runs IIa and IIb.  Between
these two epochs, substantial upgrades were made to the detector, including the addition
of a new radiation-hard silicon strip detector close to the beam line and upgrades to
the trigger system.  The instantaneous luminosity of the Tevatron increased substantially
between Run IIa and Run IIb.  The integrated luminosities are 1.0 and 8.7 fb$^{-1}$ 
for the two epochs respectively.

\section{\label{sec-trigger}Trigger }

The $\mtjj$ data were collected using all triggers
employed in D0.  
We reject events in which a muon candidate points toward the region
of impaired coverage due to the detector supports.
The trigger efficiency was determined in two steps.  In the first step
the efficiency for a suite of single muon triggers was measured using
a tag and probe analysis of 
a sample of $Z\rightarrow \mu\mu$ events and found to be about 65\%
for that sample.  
In Run IIa, the single muon triggers require the muon transverse momentum $p_T^\mu > 12$ GeV
and $|\eta_d^\mu|<2.0$, where $\eta_d^\mu$ is the pseudorapidity calculated assuming the muon originated
at the center of the detector.
Owing to the higher instanteous luminosity in Run IIb, the trigger
requirements were tightened to $p_T^\mu>15$ GeV and $|\eta_d^\mu|<1.6$.
The single muon trigger efficiency is
parameterized as a function of the 
$\eta_d$ and $\phi$ of the muon and instantaneous luminosity (as well as
$p_T^\mu$ in Run IIa).
The background and signal events simulated by Monte Carlo (MC)
are weighted by these efficiency functions.

In the second step, we measure the ratio, $\cal R$$_{\rm all}$, of the $\ltjj$
signal sample data events collected with all triggers
to those with single muon triggers, after subtracting the expected multijet component
from both. (The signal and multijet samples are discussed in 
Sections~\ref{sec-evtsel} and \ref{sec-mj}.)  
We examine the dependences of this ratio upon the $p_T$ and $\eta$ of the $\mu$, $\tau_h$
and leading (highest $p_T$) jet.    No significant dependences
are observed, and a constant $\cal R$$_{\rm all}$ is used as an additional weighting factor for
the efficiency of the MC samples.
The use of this inclusive trigger approach gives an increase in the data sample of about
40\% compared to that from the single muon triggers alone.

For the $\etjj$ analysis, we employ a set of triggers that require 
an EM object and a jet.  
The efficiency of the electron components of 
these triggers is obtained from a tag and probe analysis of $Z\rightarrow ee$
events and is parametrized in terms of electron $p_T$ and $\eta_d$.
The efficiency of the jet component is measured in events
selected by a single muon trigger in which a jet is reconstructed offline;
the jet trigger term efficiency is then determined as a function of jet $p_T$ and $\eta_d$
on the basis of whether or not the corresponding muon plus jet trigger
condition is satisfied.
The impact of the correlation between electron and jet portions of the trigger is small.
The trigger efficiency is about 85\% for the signal processes.  

\section{\label{sec-mcsamples}Background and signal samples}

The major backgrounds for the Higgs boson search are $Z$ and $W$ bosons produced in
association with jets, $\ttbar$, and QCD 
multijet production (MJ) in which a jet simulates a lepton or hadronically decaying tau.  
Smaller backgrounds arise from boson ($W, Z$ or $\gamma$) pair production (``diboson'')
and single top quark production which is included with the $\ttbar$ background.
All but the MJ background are simulated using 
MC event generator programs and normalized
to the highest order theoretical calculations available.  
These are referred to below as ``SM'' backgrounds.
The MC simulations use
the CTEQ6L1 parton distribution functions (PDF)~\cite{cteq}.  

The $\zj$ and $\wj$ MC 
event samples are generated by {\footnotesize ALPGEN}~\cite{alpgen}, interfaced
to {\footnotesize PYTHIA}~\cite{pythia} to provide initial and final state radiation and the 
hadronization of the produced partons. The $p_T^Z$ distribution is reweighted to agree
with the D0 measurement~\cite{d0zpt}.  The $p_T^W$ is also reweighted 
using the same
experimental input, corrected for the theoretical differences 
between $W$ and $Z$ bosons expected in next-to-next-to-leading
order (NNLO) QCD~\cite{nnlowpt}. 
The $\zj$ and $\wj$ cross sections are normalized using the calculations of
Ref.~\cite{nnlovjxs} and the MSTW 2008 NNLO PDFs ~\cite{mrst2008}.

We simulate $\ttbar$ and single top quark events using the {\footnotesize ALPGEN} 
and {\footnotesize SINGLETOP}~\cite{comphep}
generators respectively, with the parton hadronization provided by {\footnotesize PYTHIA}.
The normalizations are based on 
approximate NNLO QCD calculations~\cite{topxs}.  The diboson events are generated
by {\footnotesize PYTHIA} and normalized with {\footnotesize MCFM}~\cite{mcfm}.

Higgs boson production is simulated using {\footnotesize PYTHIA}, with normalizations
taken from Ref.~\cite{tev-ichep-combo}.  The Higgs boson decays are simulated using 
{\footnotesize HDECAY}~\cite{hdecay}, and
the $\tau$ decays are obtained from {\footnotesize TAUOLA}~\cite{tauola}.  

The MC signal and background events are passed through the 
{\footnotesize GEANT3}-based ~\cite{geant} simulation 
of the detector response.  Prior to reconstructing the MC
events with the standard data programs, 
we superimpose events from a library of data events collected
from random beam crossings to account for detector noise and pileup from additional 
$p\overline p$ collisions in the same or previous bunch crossings.
The difference between the luminosity distribution
for the random events and our data sample is encoded in a weight factor applied to the MC 
events.  
Simulated events are also weighted to
account for the differences between MC and data for 
the lepton, tau, and jet identification efficiencies and for the energy scale and resolution of jets,
in addition to the trigger weights discussed in Section~\ref{sec-trigger}.

\section{\label{sec-object}Object selection criteria}

Muon candidates are required to have hits in the muon chambers before 
and after the toroidal magnets, and to be matched to a track in the 
tracking system.
Muons must be isolated from additional energy deposits 
in both the calorimeter and the tracking system. 
We require the calorimeter transverse energy, $E_T^{\rm iso}$, in the annular cone
$0.1 < R <0.4$ around the muon to be less than 2.5 GeV, 
where $R =\sqrt{\Delta\eta^2 + \Delta\phi^2}$, and require that 
the sum of the transverse momenta 
of tracks within a cone  $ R <0.5$, excluding that of the 
candidate muon, be less than 2.5 GeV.   
We reject cosmic ray induced muon candidates 
by requiring that 
a time of arrival at the muon system
scintillation counters is within 10 ns of that expected for collision products.
The muon isolation selections are reversed for a special MJ control sample discussed 
in Section~\ref{sec-mj}.

Electrons are identified using information from the EM and hadronic calorimeters,
tracking detectors and the preshower detectors to form a combined electron identification
variable, $\zeta$.  The main component of $\zeta$ is a likelihood variable, $\cal L$$_e$,
defined using eight individual variables: 
the $\chi^2$ for the transverse and longitudinal shower profile to conform 
    to that expected for an EM shower; 
the fraction of calorimeter energy observed in the EM layers; 
the number of CPS strips hit;
the $\chi^2$ of the track match to the calorimeter cluster centroid; 
the ratio of track momentum and calorimeter cluster energy; 
the number of tracks in a cone  $R <0.05$ around the electron; 
the sum of the transverse momenta of tracks within $ R <0.4$ of the candidate track;
and the distance of closest approach of the track to the primary vertex (PV).
The rms width of the calorimeter cluster and the isolation of the calorimeter
cluster from nearby energy also contribute to the determination of $\zeta$.
We require $\zeta$ to exceed a threshold that is different for electrons
in the CC and EC.
The identification 
efficiencies are parametrized in terms of $p_T$, $\eta_d$ and $\phi$.
At $p_T = 25$ GeV, the efficiencies for CC electrons are about 83\% 
and are about 50\% for EC electrons, as measured in $Z\rightarrow ee$ events.  
In addition, electron candidate tracks in the CC region are required to impinge 
upon a calorimeter module 
within the central 80\% of its azimuthal range.
The $\cal L$$_e$ variable requirement 
is modified for the MJ control sample described in Section~\ref{sec-mj}.


\begin{figure*}[t]
\begin{center}
\includegraphics[width=0.330\textwidth]{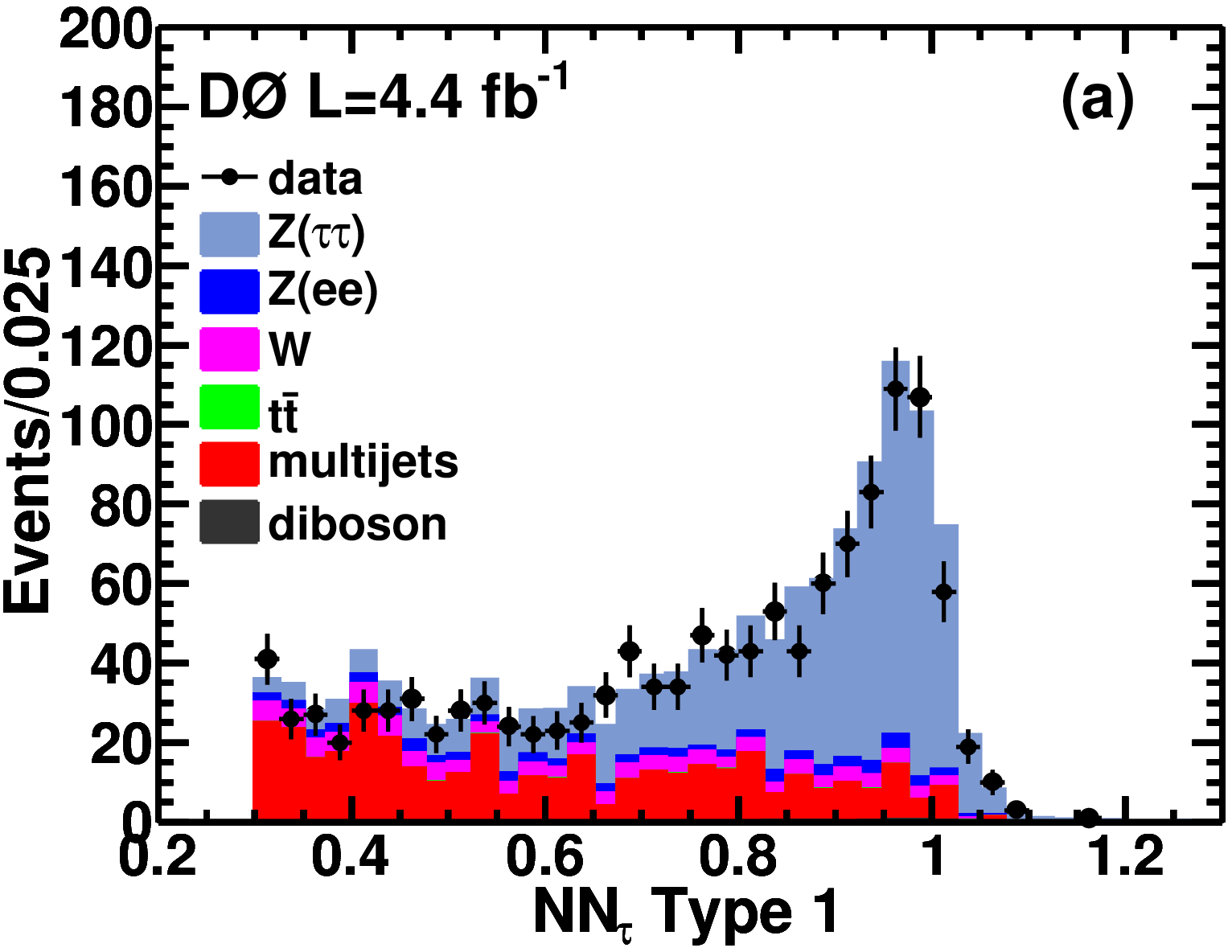}
\includegraphics[width=0.330\textwidth]{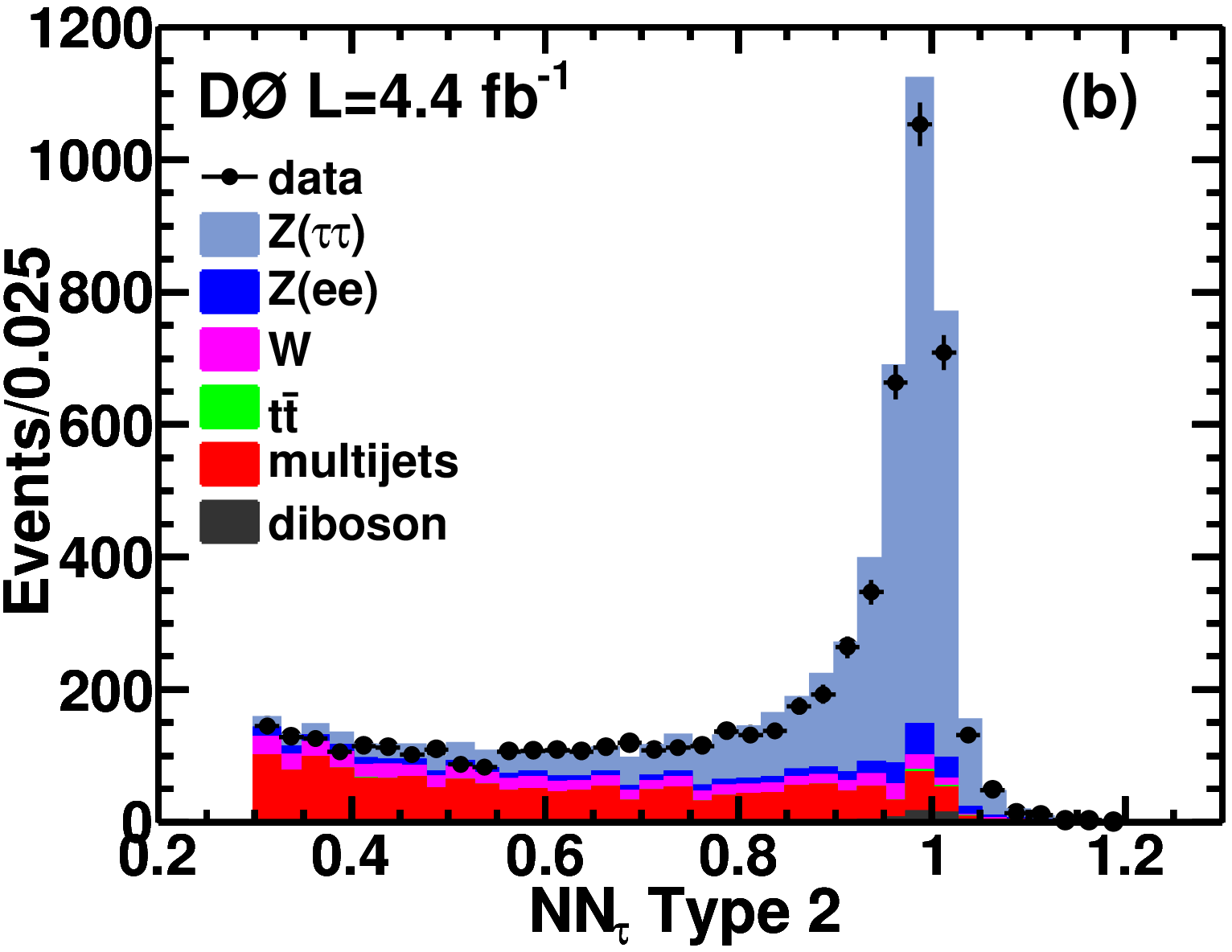} \\
\includegraphics[width=0.330\textwidth]{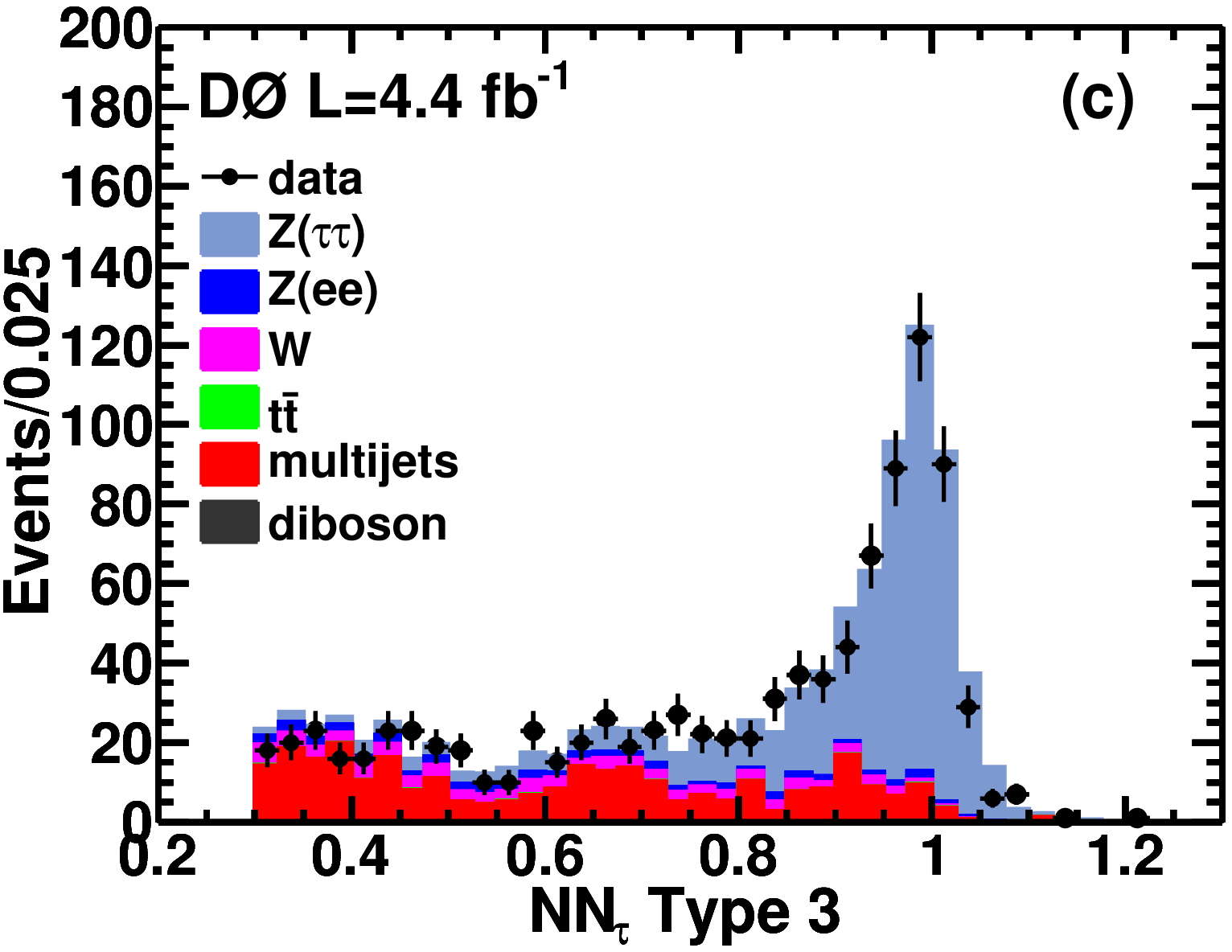}
\includegraphics[width=0.330\textwidth]{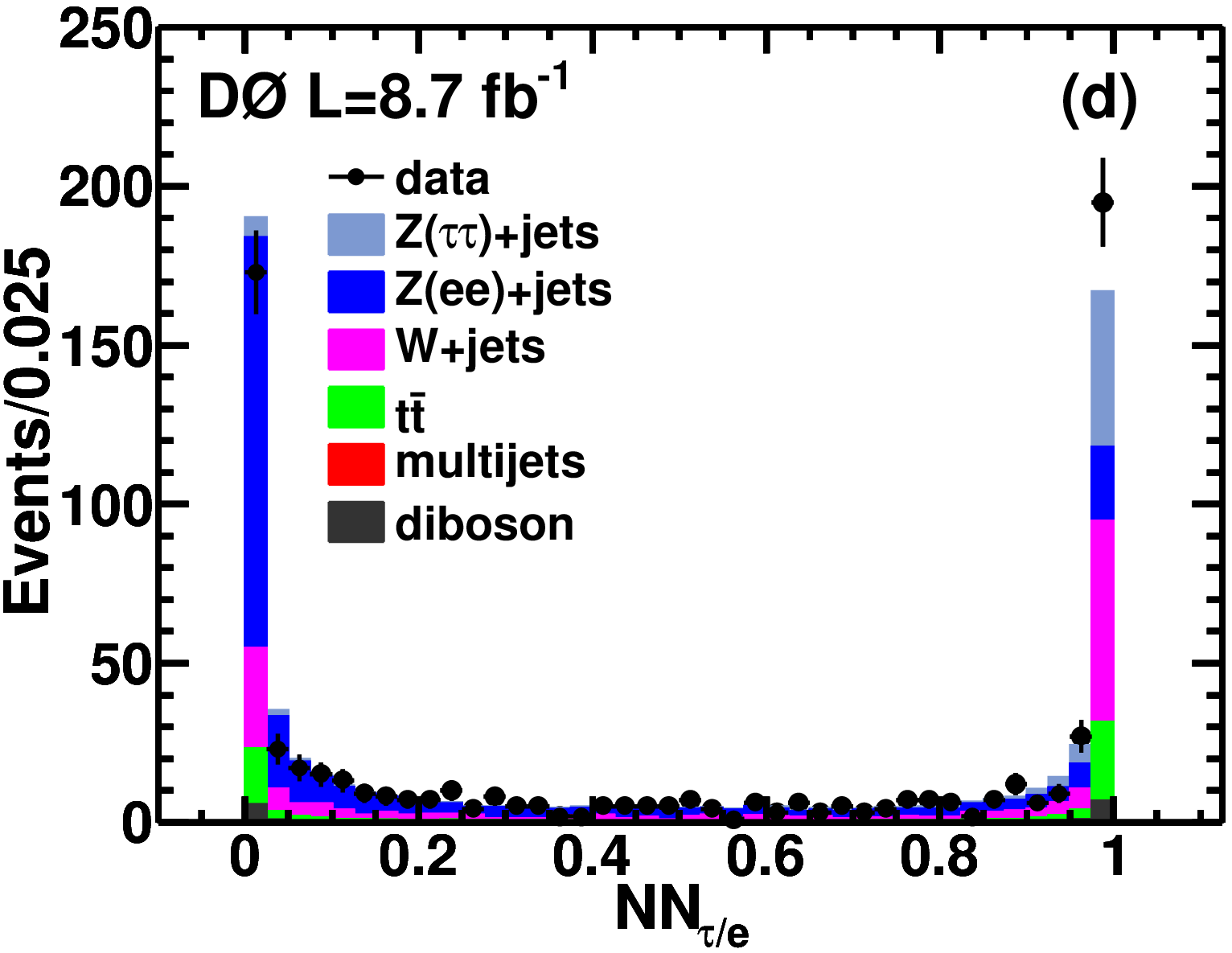} \\
\caption{\label{fig-taunn}
(color online) 
The NN$_\tau$ distribution for (a) type 1 taus, (b) type 2 taus, 
(c) type 3 taus, using data taken from
a sample of $Z\rightarrow \tau\tau$ events with no requirements
on the number of associated jets,  and (d) the NN$_{\tau/e}$ distribution for
tau type 2 from a sample of $\etjj$ events.   
}
\end{center}
\end{figure*}


We select three types of hadronically decaying tau leptons based on the number of 
tracks within a cone $R <0.3$ and the number of EM subclusters found in
the calorimeter using a nearest neighbor algorithm.  Type 1 
requires one track and no EM subclusters.  
Type 2  requires one track and at
least one EM subcluster.
Type 3  requires at least two
tracks with or without EM subclusters.  Type 3 candidates 
with exactly two tracks of opposite charge sign, for which the tau charge is ambiguous, are rejected. 
The visible $\tau_h$ transverse
energy, $E_T^\tau$, is constructed from 
the track momenta, augmented with EM calorimeter information. 
We require the sum of the track momenta associated with the $\tau_h$ ($p_T^\tau$)
to exceed (7, 5, 7) GeV, 
$E_T^\tau > (12.5, 12.5, 15)$ GeV, and 
$(p_T^\tau / E_T^\tau) > (0.65, 0.5, 0.5)$ for tau types (1, 2, 3) respectively, and
the highest $p_T$ track for type 3 taus must exceed 5 GeV.
We construct a neural network, NN$_\tau$~\cite{taunn}, based on energy deposition patterns and isolation
criteria in the calorimeter and tracking systems for each tau type to discriminate a 
tau from a misidentified jet.  Figure~\ref{fig-taunn} shows illustrative NN$_\tau$ distributions
taken from a control data sample of $Z\rightarrow\tau\tau$ candidates.
We require NN$_\tau$ to be greater than (0.92, 0.90, 0.91) for
tau types (1, 2, 3) respectively.
For type 2 taus, we construct a second neural network, NN$_{\tau/e}$, to differentiate 
taus from electrons.  The distribution of NN$_{\tau/e}$, taken from the
$\etjj$ sample described in Section~\ref{sec-evtsel}, 
is shown in Fig.~\ref{fig-taunn}(d).  We select events with
NN$_{\tau/e} > 0.5$.

Jets are reconstructed using an iterative midpoint cone algorithm~\cite{jetalg} with a cone size
$R = 0.5$.  
We require at least two tracks associated with the jet that point to the PV (vertex 
confirmation) in Run IIb
due to the higher multiplicity of collisions within a bunch crossing.
Jet energies are corrected to the particle level for out-of-cone showering, underlying event 
energy deposits and pileup from neighboring beam crossings, 
and for the effects of energy carried by muons and neutrinos 
when there is evidence for semileptonic decays of the jet particles.   Jets in data are corrected 
for energy scale and resolution using $\gamma$+jet and dijet samples.
The MC jets are corrected for energy scale and resolution, as well as for the jet identification
efficiency to bring the MC responses into agreement with data.
For the  MC samples rich in quark jets
($t\overline t$ and diboson), there is an additional calibration
applied to the jet energy that accounts for
the differences between the responses of quark jets and the dominantly gluon jets for which the 
jet energy scale correction was obtained.

The missing transverse energy, 
$\met$, is computed from the observed transverse energy deposits in the calorimeter 
and is adjusted for 
the appropriate energy scale corrections for all objects, for isolated muons that
deposit less than their full energy in the calorimeter, and for the unclustered energy in the 
calorimeters not associated with jets or EM objects.
We define a quantity $\cal S$ that measures the significance of the $\met$ to be
different from zero, based on the measured resolutions of the components of the
$\met$ calculation~\cite{metsig}.

\section{\label{sec-evtsel}Event selection criteria}

We select a sample of events (``signal sample'') with the criteria given below.  Some
of these differ for the Run IIa and Run IIb selection owing to the differences in the detector, triggers
and luminosity, and the fact that the jet vertex confirmation was not applied in Run IIa, leading
to differences in the modelling of jet related variables.

For the $\mtjj$ selection we require:

\begin{itemize}
\item
a muon as defined in Section~\ref{sec-object} with $p_T^\mu>12$ (15) GeV
and $|\eta_d^\mu|<2.0$ (1.6), for Run IIa (Run IIb);
\item
at least one hadronic tau as defined in Section~\ref{sec-object} 
with $|\eta_d^\tau| \leq 2$.  The tau candidate with the highest $p_T$ is chosen
for the analysis, and must have a charge sign that is opposite to the muon;
\item
two jets with $|\eta_d^{\rm jet}|<3.4$; the leading jet (``jet1'') is required to have
$p_T^{\rm jet1}>20$ GeV and a second-leading jet (``jet2'') to have $p_T^{\rm jet2}>15$ GeV;
\item
no other electron with $p_T^e > 10$ GeV  and
no other muon with $p_T^\mu>10$ GeV 
to retain orthogonality to other D0 searches
for the SM Higgs boson;
\item
the scalar sum of all jet $p_T$'s in the event ($H_T$) must be greater than 80 GeV for Run IIa
to improve the modelling of jet related variables.
\end{itemize}

For the $\etjj$ selection we require:

\begin{itemize}
\item
an electron as defined in Section~\ref{sec-object} with $p_T^e>15$ GeV,
and $|\eta_e|<1.1$ or $1.5<|\eta_e|<2.5$;
\item
at least one hadronic tau as defined in Section~\ref{sec-object} 
with $|\eta_d^\tau| \leq 2$. The tau with the highest $p_T$ is chosen
for the analysis, and must have a charge sign that is opposite to the electron;
\item
no type 1 taus with $1.05<|\eta|<1.5$, and type 2 taus to have
NN$_{\tau/e}> 0.5$, 
to reduce the $Z(ee)+$jets background in which an electron is misidentified as a tau;
\item
two jets with $|\eta_d^{\rm jet}|<3.4$; 
$p_T^{\rm jet1}>25$ (20) GeV for Run IIa (Run IIb) 
and $p_T^{\rm jet2}>15$ GeV;
\item
no other electron with $p_T^e > 12.5$ GeV and
no muon with $p_T^\mu>12$ GeV and $|\eta_d^\mu|<2.0$ for orthogonality to other D0 
SM Higgs boson searches;
\item
the $\met$ significance variable is required to be $\cal S$ $>3$ (2) for Run IIa (Run IIb) to
reduce the MJ and $Z\rightarrow ee$ backgrounds.
\end{itemize}

Events with the same selection criteria except that the $\ell$ and $\tau_h$ have
the same charge sign are retained for estimating the MJ background in both
the $\mtjj$ and $\etjj$ analyses (SS signal samples).

\section{\label{sec-mj} Multijet background}

The MJ background arising from misidentification of leptons or taus
by the detector reconstruction algorithms is difficult to simulate, so 
it is estimated using data.  We define a sample of MJ-enriched events (MJ control sample, $\cal M$)
which is large compared
with the size of the signal sample. 
We use the events in $\cal M$, after subtraction of
the small residual SM backgrounds simulated by MC, to provide the shapes of the MJ background
kinematic distributions.  
We also use the $\cal M$ sample to obtain the MJ background normalization.
For each tau type, the SM subtracted sample $\cal M$ is divided   
into the opposite sign $\ell\tau_h$ (OS) and same sign $\ell\tau_h$ (SS) samples.
The ratio of OS and SS events in $\cal M$ is used to scale the 
number of MJ events in the SS signal sample to obtain the normalization
for the MJ background in the signal sample.
The MJ yield in the SS
signal sample is obtained after subtracting the small SM backgrounds.  
Denoting the number of events in $\cal M$ by $M$ and the number
of events in the SS or OS signal samples by $N$, the method is expressed by:

$$ N_{\rm OS}^{\rm MJ} = \rho(N_{\rm SS}^{\rm data}-N_{\rm SS}^{\rm SM}) $$

with 

$$ \rho = (M_{\rm OS}^{\rm data} - M_{\rm OS}^{\rm SM})/
   (M_{\rm SS}^{\rm data} - M_{\rm SS}^{\rm SM})~~.  $$

This background estimate is computed separately for each tau type and summed
to give the total MJ background.   

For the $\mtjj$ analysis, the MJ sample $\cal M$ is obtained by reversing at least
one of the muon isolation requirements.  The tau selection requirements are the same
as for the signal sample.
The MJ purity in this sample is about 97\%.
We observe no significant dependence of the $\rho$ factors on the $p_T$ or $\eta$
of $\mu$, $\tau_h$ or jets, and therefore take them to be constant.  Their values are
within about 15\% of unity and are similar for the three tau types, and
in Run IIa and Run IIb.  We observe that the modelling of the shapes of variables
which employ the jet $p_T$'s needs improvement and so we adopt a reweighting 
for the MJ events based upon the comparison of the distribution shapes 
in the SS signal sample and the MJ control sample.  We find that the simple
reweighting function $A e^{-B {H_T}}$, fitted to the ratio of the $H_T$ distributions for
the SS signal and MJ control
samples, gives adequate modelling.

For the $\etjj$ analysis, $\cal M$ is obtained by requiring the tau selection
$0.3 <{\rm NN}_\tau < 0.9$, and by placing an upper bound of 0.85 on the $\cal L$$_e$
in the electron identification. In forming $\cal M$,
no requirement is made on the ~$\met$ significance.
$\cal S$.  For the Run IIb data, the cut on $\cal S$ in the 
signal selection leaves a
negligibly small MJ contribution in the SS signal sample after the SM background subtraction.  
Owing to the absence of jet vertex confirmation,
some MJ background remains in Run IIa and the procedure used in the $\mtjj$ analysis 
for its estimation is followed, except that no $H_T$ reweighting is needed.  
For Run IIa, the MJ purity in $\cal M$ is 98\%
and the $\rho$ values are about 1.25.

\section{\label{sec-yields}Event Yields}

The numbers of data and expected background events are given in
Table~\ref{tab-yield} for the $\mtjj$ and $\etjj$ analyses.
Taking into account the systematic uncertainties on the backgrounds 
(see Section~\ref{sec-syst})
and uncertainties on the MC statistics, the predicted backgrounds are in 
agreement with the observed yields.
The $\etjj$ yields are smaller than those for $\mtjj$ primarily
because of the requirements made for the $\etjj$ analysis on $\cal S$, $\eta_\tau$, 
$\phi_e$ in the CC region, and NN$_{\tau/e}$ to reduce the
MJ and $Z\rightarrow ee$ backgrounds.
Representative expected signal yields for the nine production
and decay processes are given in Table~\ref{tab-signals}.


\begin{table}[htbp]
\caption{\label{tab-yield}
For each analysis channel, 
the number of  background events expected from SM processes, MJ background, 
and observed in data, for individual and sum of all tau types   
after preselection. 
``type'' denotes $\tau$ type, 
``$V$j'' denotes $W$ or $Z$ + jets and ``DB'' denotes diboson processes. The
uncertainty on the sum of all backgrounds includes both MC statistical
and systematic uncertainties.
}

\begin{tabular}{crrrrrrD{,}{\pm}{-1}r} \hline \hline
type & $\ttbar$~~ & ~$W$j & $Z_{\ell \ell}$~j & $Z_{\tau\tau}$~j & DB 
    & MJ & \multicolumn{1}{c}{$\Sigma$~{\rm Bkd}}  & Data  \\ \hline 

\multicolumn{9}{c}{$\mtjj$ analysis} \\ \hline
 1    &  15.3 & 10.2  & 4.4  & 37.1  & 2.3  & 39.1  & 108.4,7.4    & 119  \\ 
 2    & 121.3 & 65.2  & 29.3 & 241.8 & 14.5 & 135.4 & 607.5,47.1   & 684  \\ 
 3    &  20.2 & 39.1  & 4.4  & 54.5  & 3.2  & 50.6  & 172.1,12.4   & 187  \\ 
 All  & 156.9 & 114.5 & 38.1 & 333.4 & 16.0 & 225.1 & 888.0,49.2   & 990  \\ \hline

\multicolumn{9}{c}{$\etjj$ analysis}  \\ \hline
  1      &  4.5   &  4.6   &  0.0 & 9.8     & 0.9  & 3.1  &  23.0,1.8  &  15  \\
  2      &  57.7  &  64.9  & 66.6  & 91.7   & 8.3  & 1.7  & 290.8,21.0 &  261  \\
  3      &  27.2  &  47.2  &  2.4  & 28.5   & 3.7  & 14.6 & 123.7,9.6  &  124  \\ 
  All    & 89.4   &  116.7 & 69.1  & 130.0  & 12.9 & 19.4 & 437.5,23.2 &  400  \\ \hline \hline
\end{tabular}
\end{table}

\begin{table*}[htbp]
\caption{\label{tab-signals}
For each analysis channel, 
the number of  signal events expected for each of the nine production and decay processes,
prior to the separation into the T and W subsamples discussed in the text.
}

\begin{tabular}{crrrrrrrrrr} \hline \hline
~~~~$M_H$~~ & ~~~~~~HZ & ~~~~ZH$_{\tau\tau}$ & ~~~~WH$_{\tau\tau}$ & 
~~GF$_{\tau\tau}$ & ~~~~VBF$_{\tau\tau}$ & 
~~~~ZH$_{WW}$ & ~~~~WH$_{WW}$ & ~~~~GF$_{WW}$ & ~~~~VBF$_{WW}$ & ~~~~Total \\ \hline 

\multicolumn{11}{c}{$\mtjj$ analysis} \\ \hline
105 & 0.19 & 0.47 & 0.71 & 0.66 & 0.37 & 0.03 & 0.04 & 0.10 & 0.01 & 2.58 \\
115 & 0.15 & 0.38 & 0.57 & 0.53 & 0.34 & 0.09 & 0.15 & 0.08 & 0.05 & 2.34 \\
125 & 0.10 & 0.27 & 0.40 & 0.45 & 0.26 & 0.19 & 0.37 & 0.18 & 0.14 & 2.35 \\
135 & 0.06 & 0.16 & 0.23 & 0.30 & 0.18 & 0.33 & 0.63 & 0.31 & 0.24 & 2.43 \\ 
145 & 0.03 & 0.08 & 0.11 & 0.17 & 0.10 & 0.47 & 0.81 & 0.59 & 0.37 & 2.72 \\  \hline

\multicolumn{11}{c}{$\etjj$ analysis} \\ \hline
105 & 0.14 & 0.34 & 0.53 & 0.32 & 0.26 & 0.01 & 0.01 & 0.02 & 0.00 & 1.63 \\
115 & 0.11 & 0.31 & 0.47 & 0.32 & 0.24 & 0.04 & 0.05 & 0.01 & 0.01 & 1.56 \\
125 & 0.07 & 0.22 & 0.33 & 0.30 & 0.21 & 0.09 & 0.14 & 0.04 & 0.04 & 1.46 \\
135 & 0.05 & 0.15 & 0.21 & 0.21 & 0.15 & 0.17 & 0.26 & 0.11 & 0.08 & 1.38 \\ 
145 & 0.02 & 0.07 & 0.11 & 0.12 & 0.08 & 0.25 & 0.35 & 0.16 & 0.11 & 1.26 \\  \hline \hline

\end{tabular}
\end{table*}

\section{\label{sec-mva}Multivariate Analysis}

The number of background events greatly exceeds the expected number of Higgs boson signal
events so we employ multivariate techniques that discriminate signal from background
by taking into account the correlations among the variables.   
The multivariate strategy for this analysis is complicated by the presence
of many distinct signals of comparable size, each with its own
characteristic kinematic properties.

We tested several artificial 
learning techniques including neural networks and decision trees using the 
TMVA suite of programs~\cite{tmva} to implement
the multivariate methods.  We find an optimum performance with the gradient boosted
decision tree classifier (BDT) which offers  
the advantage over neural networks that the use of variables that
do not discriminate significantly between a particular signal and background,
or are highly correlated with other variables, 
do not compromise the classifier performance.  
In the BDT~\cite{bdt1,bdt2} approach, a series of splittings of the event sample
is made at a sequential set of nodes into 
background-like and signal-like subsample nodes.  
The splitting is based upon the 
purity of signal and background events in a given node $N$
and its signal-like and background-like daughter nodes $S$ and $B$.
Purity is defined as $p=s/(s+b)$ where $s$ ($b$) is the number of
signal (background) events in the node.  
In the training, the event category is known either from the MC generation 
or MJ control samples.  
The optimum splitting is achieved by maximizing the decrease of the 
Gini index~\cite{gini}
$i = 2p(1-p) = 2sb/(s+b)^2$
going from the parent node $N$ to the two daughter nodes $S$ and $B$,
considering all choices of input variable and cut thresholds for that variable.
Each such subsample is subjected to further splitting until the sample sizes reach
a preset value.  At each node, a random sampling of events is chosen from the full
sample to help mitigate the effects of finite statistics.   
The training is recursive with misclassified events in one cycle being reweighted for 
the next cycle. The TMVA training is controlled by parameters such as the maximum
number of trees, the degree of reweighting in successive cycles, the fraction
of the full sample used at each node and the number of cycles allowed.   We
varied these parameters around their nominal settings to obtain optimum values for
our analysis.

\subsection{\label{sec-vars}BDT variables}

\begin{table}[htbp]
\caption{{\small  Variables used for $\mtjj$ and $\etjj$ analyses in BDT training. 
As indicated, three variables are used for only one analysis.
} 
}
\begin{center}
\begin{tabular}{l} \hline \hline 
{\it transverse momentum variables}            \\ \hline
 $p_T(\ell)$                                   \\
 $p_T(\tau)$                                   \\
 $p_T(j_1)$                                    \\
 $p_T(\ell\tau~\met)$                          \\
 $H_T$                                         \\ 
 $S_T$                                         \\
 $V_T$                                         \\ \hline
{\it angular variables}                        \\ \hline
 $\Delta \eta(j_1,j_2)$                        \\
 $\Delta R(j_1,j_2)$                           \\
 $\Delta\phi(\ell\tau,~j_1j_2)$                \\
 $\Delta \phi(\ell,~j_1)$                      \\
 $\Delta\phi(\met,~\mpt)$                      \\
 min$\Delta \phi(~\met,~ \ell{\rm~or~}\tau)$   \\
 max$\Delta \phi(~\met,~ \ell{\rm~or~}\tau)$   \\
 min$\Delta\phi(~\met, j_1{\rm~or~}j_2)$       \\
 max$\Delta\phi(~\met, j_1{\rm~or~}j_2)$       \\ 
 $\cos \hat{\theta}$                           \\ \hline
{\it mass variables}                           \\ \hline
 $M(j1~j2)$                                    \\
 $M(\tau \tau)$                                \\
 $M_{WW}$                                      \\
 $M(\ell \tau j1 j2)$                          \\
 $M_T(~\met,\ell)$~[$\etjj$ only]                          \\
 $M_T(~\met,\tau)$                             \\
 $M_T(~\met,\ell\tau)$                         \\
 min$M_T(~\met,~\ell{\rm~or~}\tau)$            \\ \hline
{\it other variables}                          \\ \hline
 $N_{\rm soln}(\tau \tau)$                     \\
 $N_{\rm soln}(WW)$                            \\ 
 $\mht$/$H_T$                                  \\
 $A(~\met,~\mht)$                              \\
 $\cal S$                                      \\
 NN$_\tau$~[$\mtjj$ only]                                     \\
 electron identification, $\zeta$ ~[$\etjj$ only]          \\ 
 $M_H$                                         \\ \hline \hline
\end{tabular}

\label{tab-bdtinputs}
\end{center}
\end{table}

We examine a large set of potentially discriminating kinematic variables with which to train the 
multivariate analyses.  
We choose a subset of well-modelled variables for which the agreement of data with
expected background is good and which discriminate between at least one individual
signal and background.
In these distributions the expected signal contribution to the data is small.
Table~\ref{tab-bdtinputs} shows the variables used for both the $\mtjj$ and $\etjj$
analyses.  

In Table~\ref{tab-bdtinputs}, 
$H_T$ is the scalar sum of all jets with $p_T>15$ GeV, 
$S_T = H_T + |p_T^\ell| + |p_T^\tau|$, and
$V_T$ is the magnitude of the vector transverse momenta of $\ell, \tau_h$ and all jets.
The variable~$\mht$ is the magnitude of the vectorial sum of all jet transverse momenta,
and ~$\mpt$ is the magnitude of the vector sum of all tracks emanating from the primary vertex.
The variable $A(~\met,~\mht)$  is the ratio of the difference and the sum of
~$\met$ and ~$\mht$.
Variables  $\Delta\phi(a,b)$ are the difference in azimuthal angle between
object $a$ and the object, or pair of objects, $b$, 
and similarly for the $\Delta\eta$ and $\Delta R$
variables.
The variable $\hat{\theta}$ is the angle between the dijet system and the proton
beam direction in the laboratory frame.
Variables $M(a~b ...c)$ are invariant masses of objects $a,b, ... c$ 
and $M_T(~\met, a)$ is the transverse mass computed from
$M_T^2 = 2 E_T^{a}~\met (1-\cos\phi)$ where $\phi$ is the azimuthal
angle between ~$\met$ and the object, or pair of objects, $a$.


\begin{figure*}[t]
\begin{center}
\includegraphics[width=0.300\textwidth]{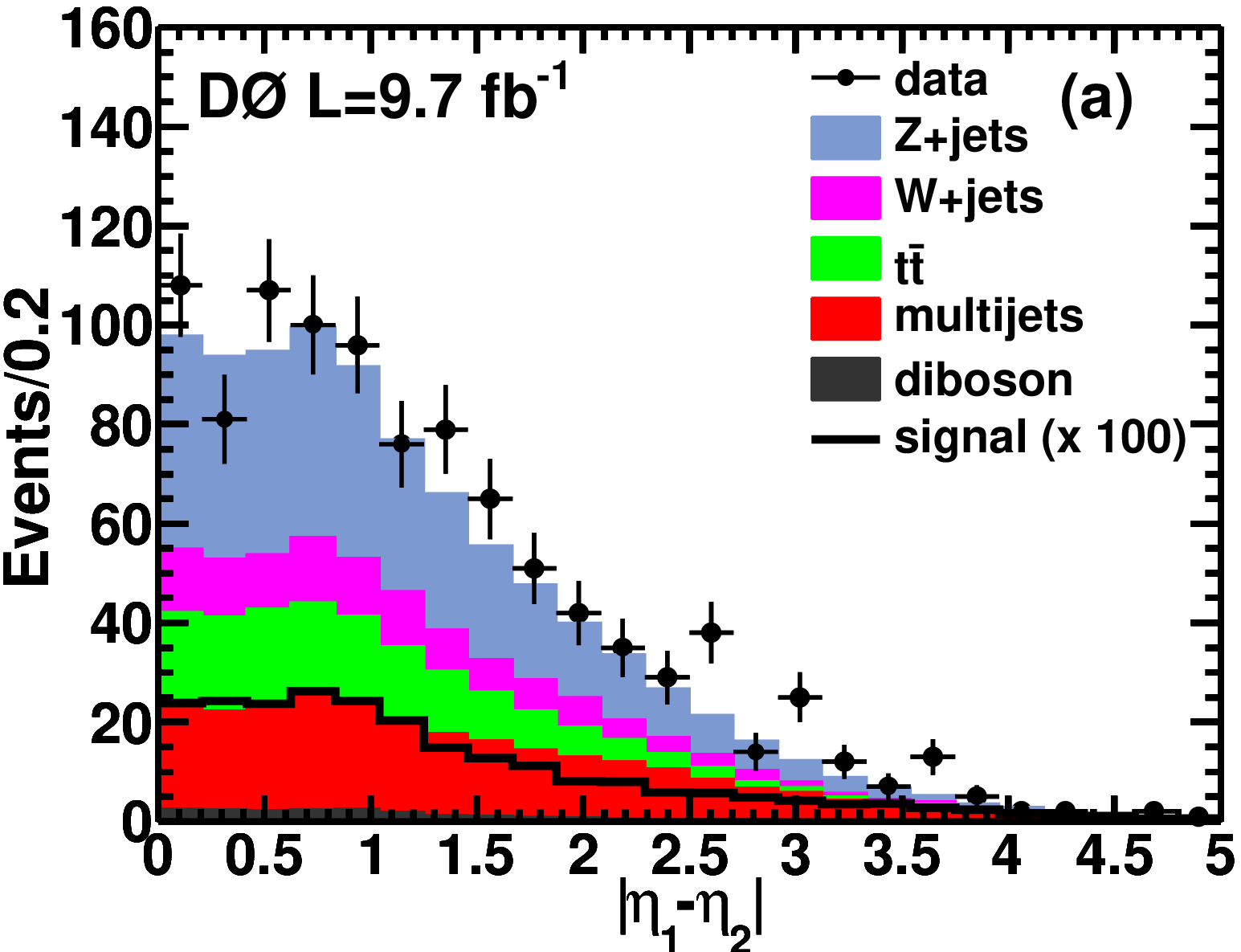}
\includegraphics[width=0.300\textwidth]{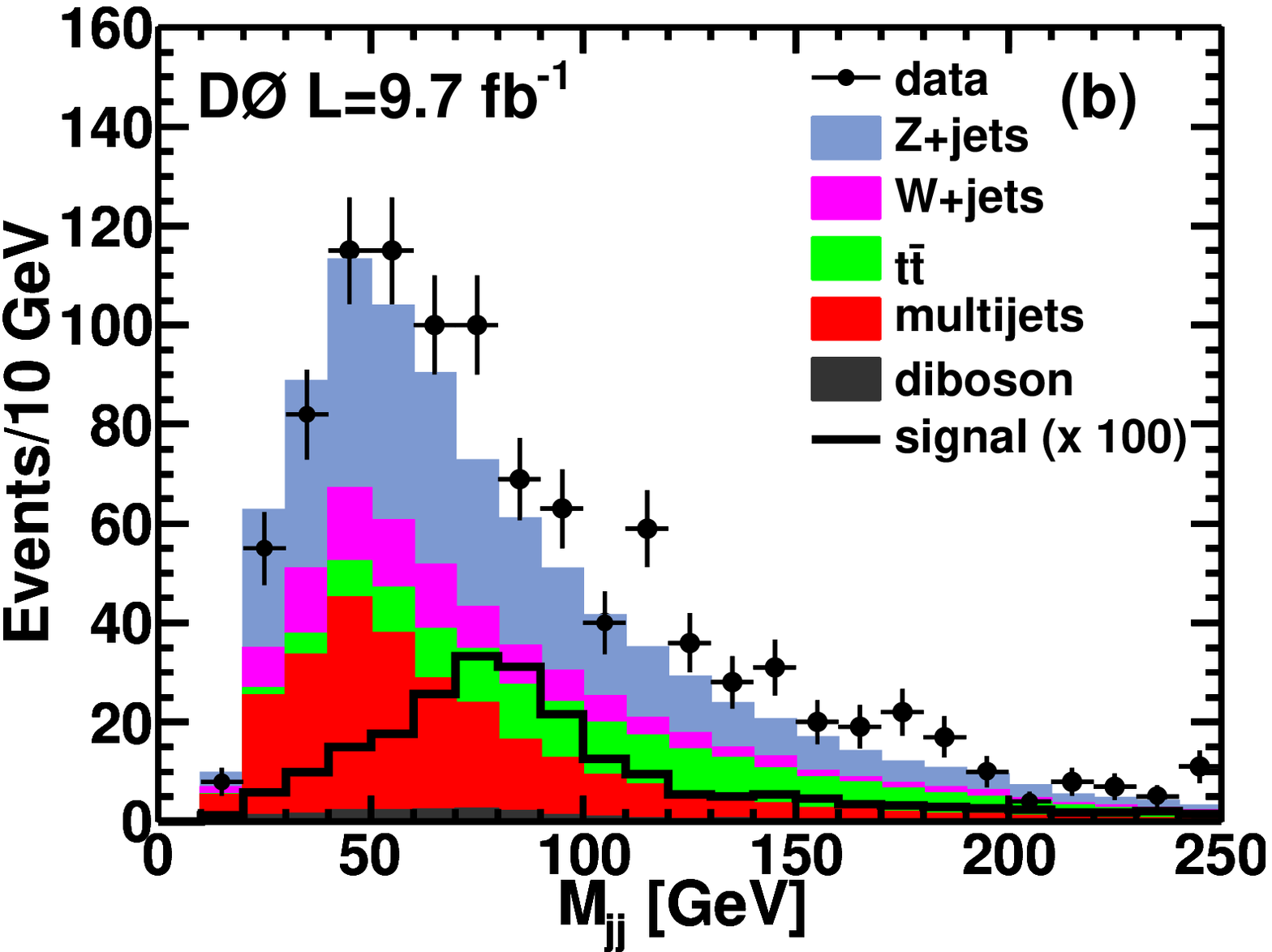} 
\includegraphics[width=0.300\textwidth]{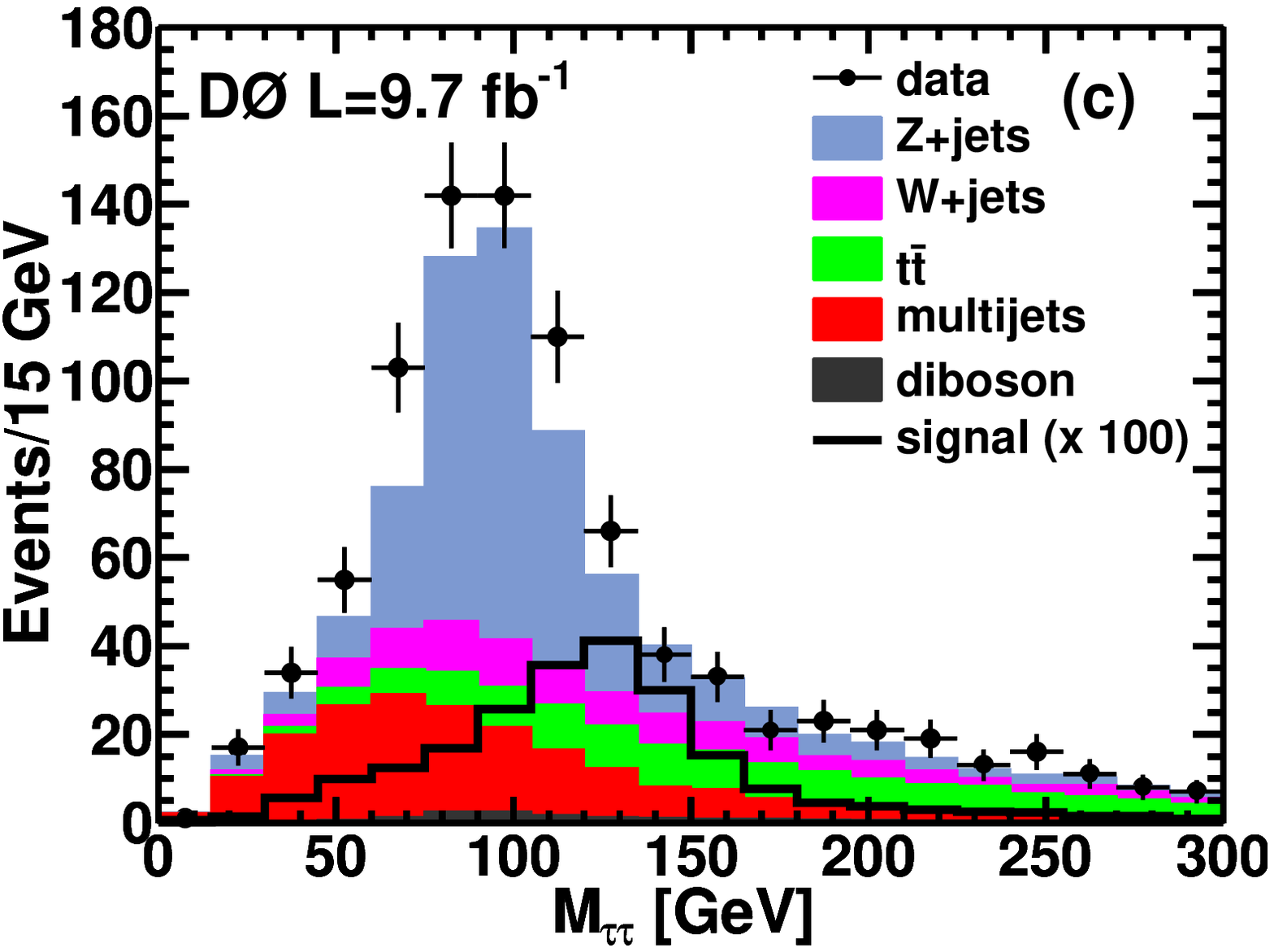} \\
\includegraphics[width=0.300\textwidth]{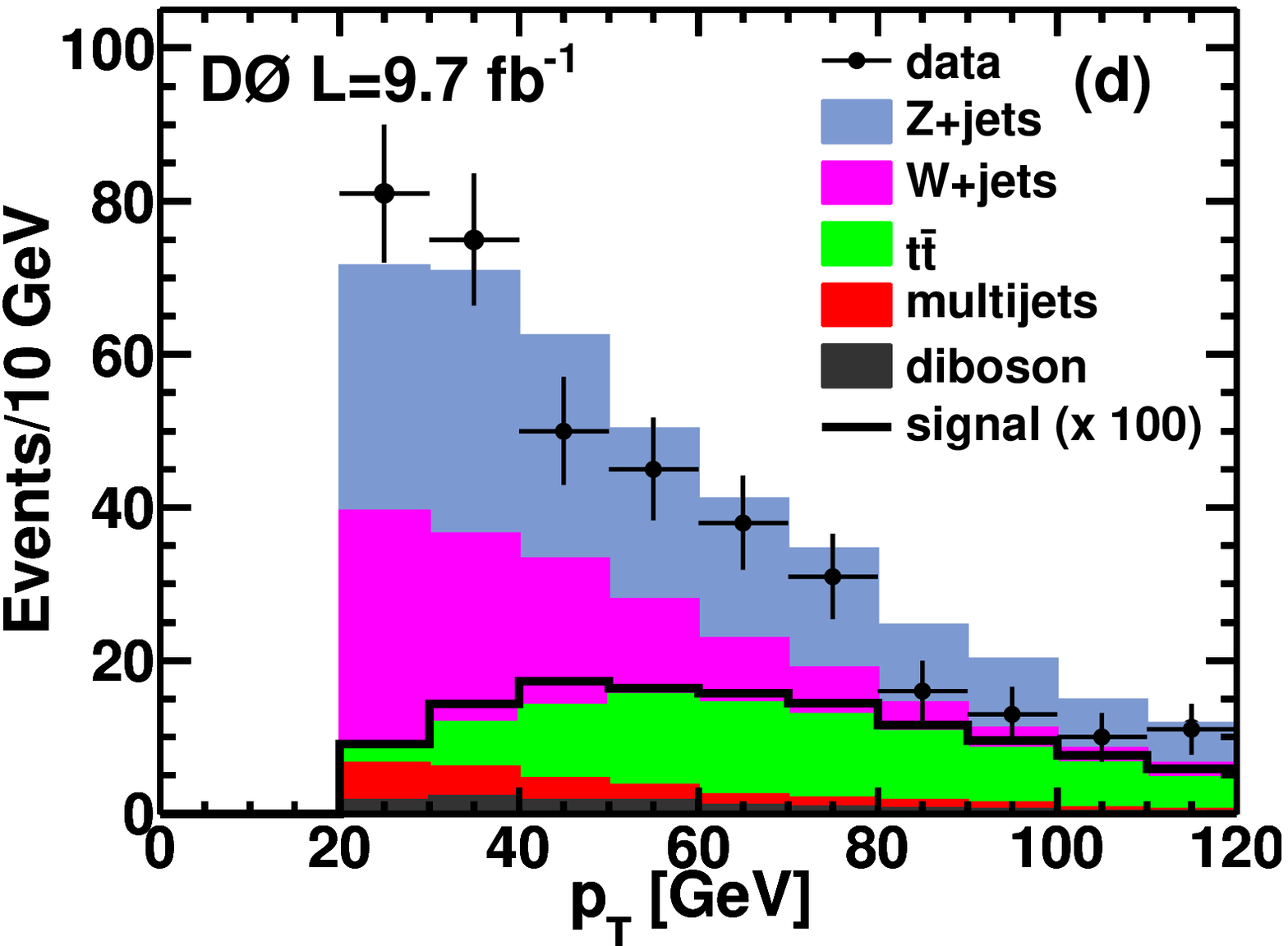} 
\includegraphics[width=0.300\textwidth]{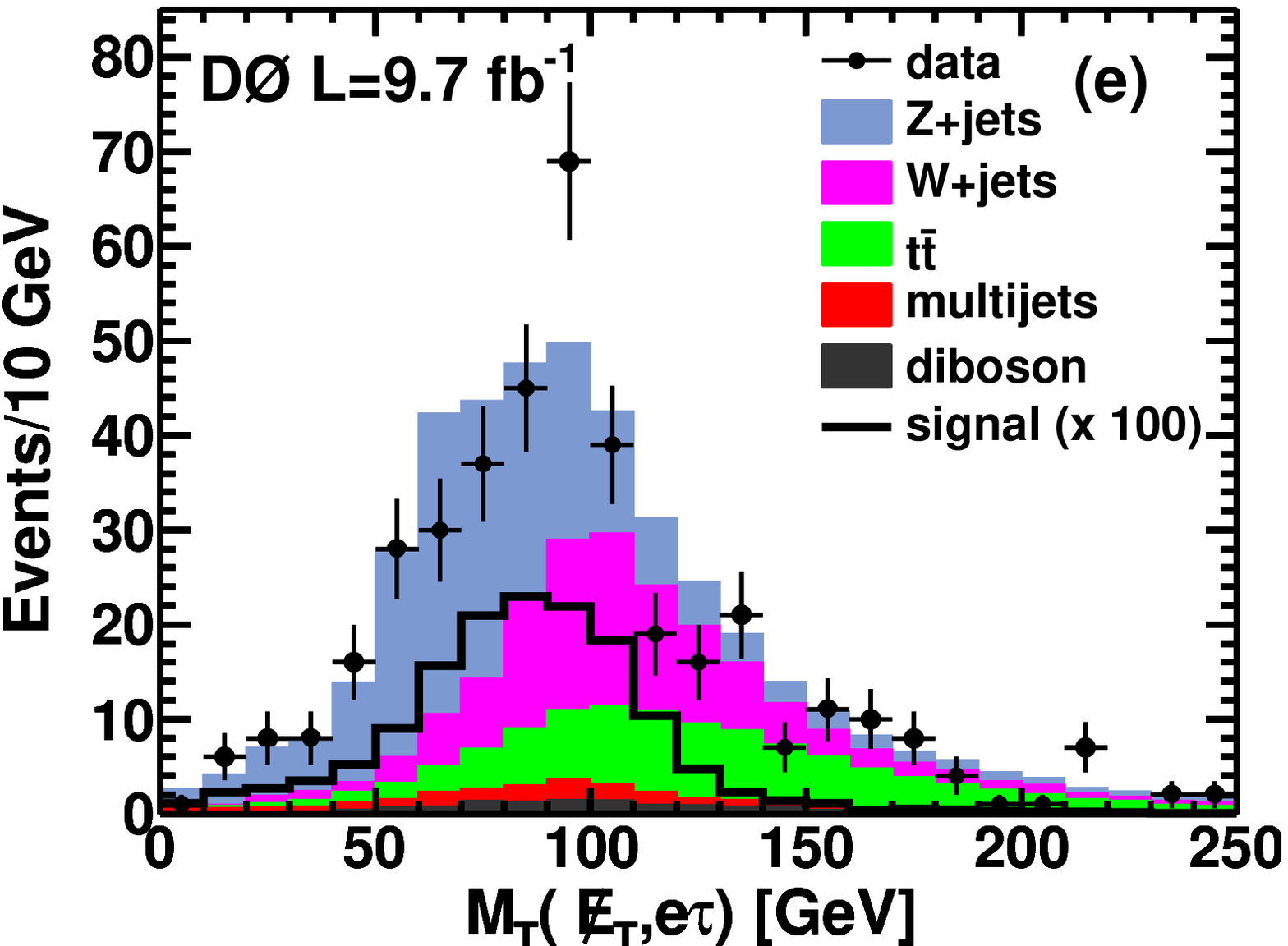}
\includegraphics[width=0.300\textwidth]{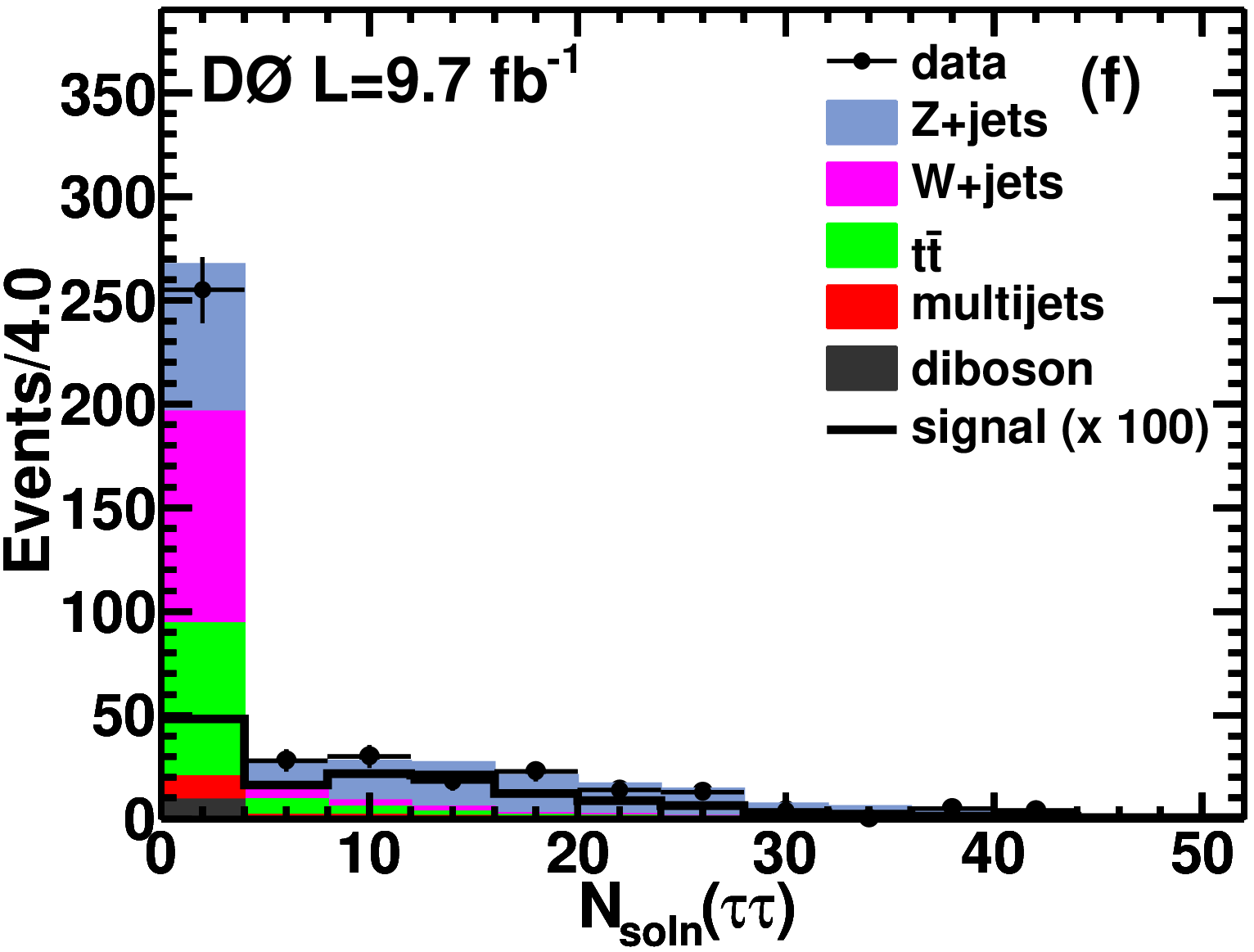} \\
\caption{\label{fig-bdtinputs}
(color online) 
Input variables for the $\mtjj$ BDT training: 
(a)  $\Delta \eta(j_1,j_2)$, (b)  $M(j1~j2)$, (c)  $M(\tau \tau)$,
and for the $\etjj$ BDT training:
(d)  $p_T(j_1)$, (e) $M_T(\met , \ell\tau$), (f)  $N_{\rm soln}(\tau\tau)$.    
}
\end{center}
\end{figure*}


The $M(\tau \tau)$ and $M(WW)$ variables are determined using
the Missing Mass Calculator (MMC) method~\cite{mmc}.  
These masses are under-constrained owing to the neutrinos
from $\tau$ or $W$ decay.
We compute the most likely $\tau\tau$ mass by scanning over a grid 
in the 3-dimensional space of
the azimuthal angle separations of the visible $\tau$ decay products and
the neutrino(s) for each $\tau$, 
and the invariant mass of the multiple neutrinos
from one of the $\tau$'s (e.g. $\tau\rightarrow\ell\nu\overline{\nu}$),
given the constraints from the measured momenta of the visible decay products and the 
known $\tau$ mass.  
At each grid point, the calculated $\tau\tau$ mass
is weighted by the probability for finding the $\Delta R$ at that point 
between the visible particles and the neutrinos.
For $H\rightarrow WW$ in the mass region considered, one $W$ is virtual.
The MMC mass is calculated for the $WW$ system taking the
mass of the virtual $W$ to be the most likely value (38 GeV) for $M_H=125$ GeV.
The variables  $N_{\rm soln}(\tau \tau)$  and  $N_{\rm soln}(WW)$  
are the number of physical mass solutions found in the grid search.
Figure~\ref{fig-bdtinputs} shows the data and background distributions of 
representative BDT input variables.

One additional variable, the mass of the
hypothesized Higgs boson, $M_H$, is used in training the BDTs as discussed
below in Section~\ref{sec-globalbdt}.

\subsection{\label{sec-tautauww}Separation of T and W subsamples}

\begin{figure}[t]
\begin{center}
\includegraphics[width=0.400\textwidth]{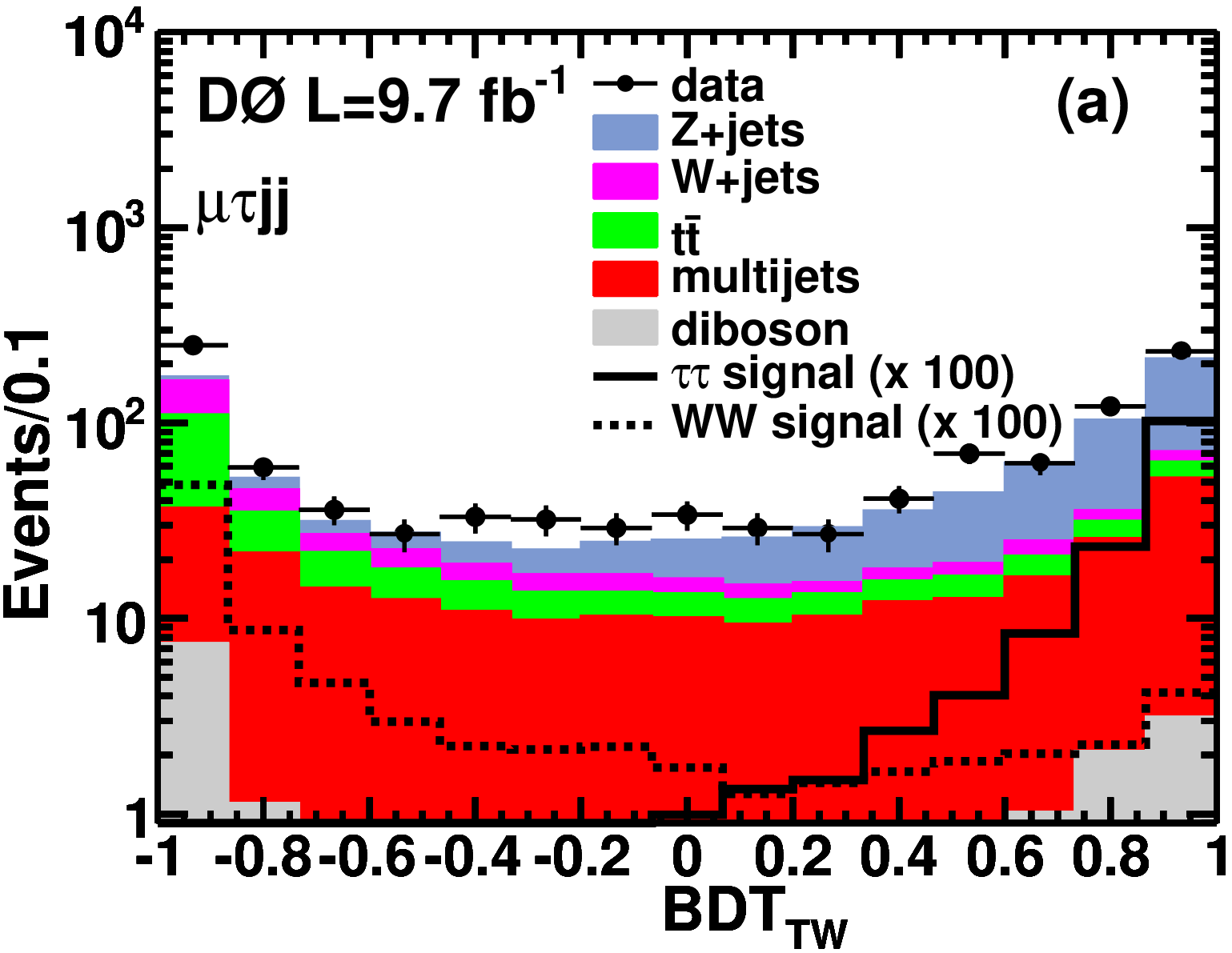} \\ 
\includegraphics[width=0.400\textwidth]{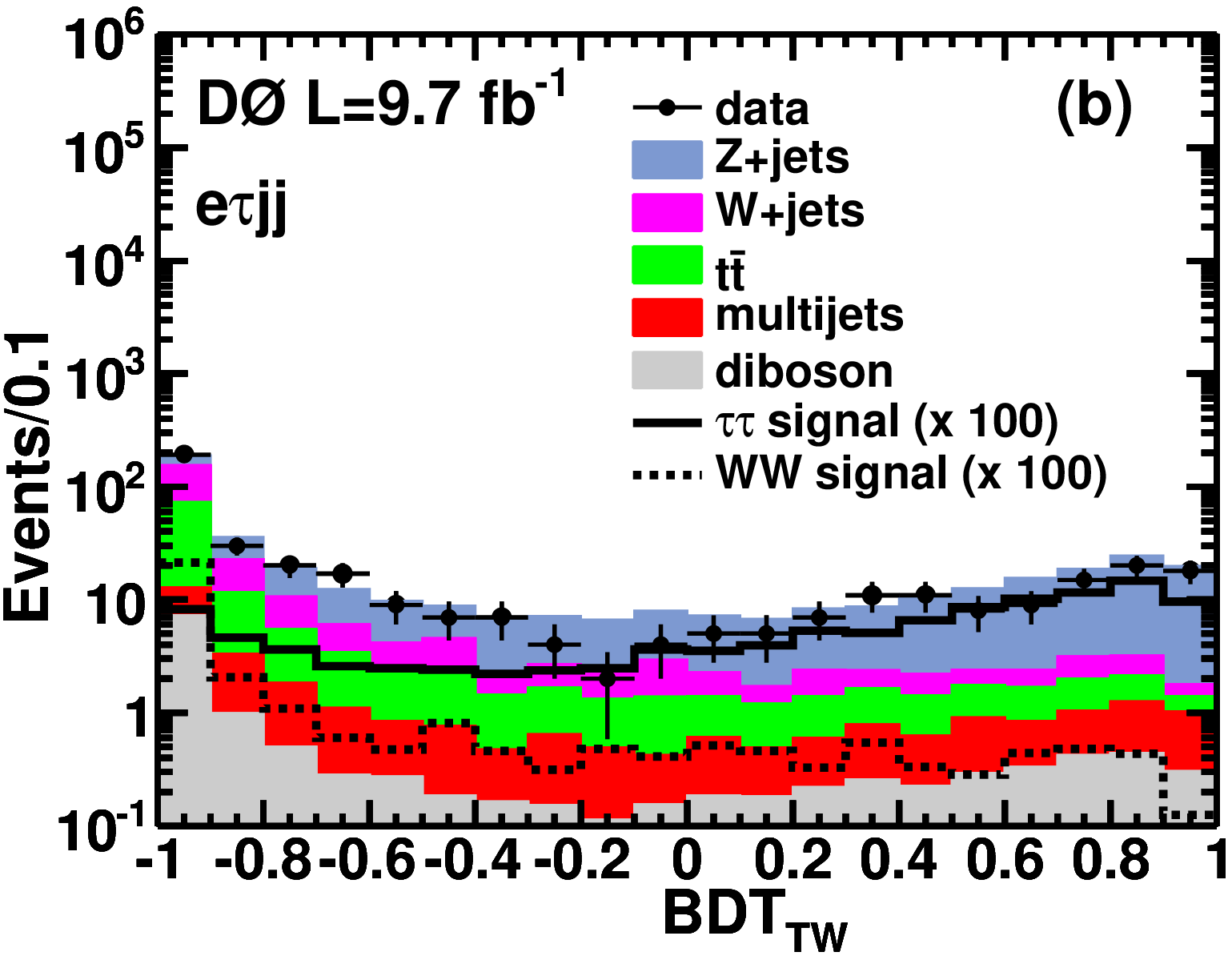} \\
\caption{\label{fig-bdttauw}
(color online) 
BDT$_{\rm TW}$ outputs from the training of $H\rightarrow\tau\tau$ vs. $H\rightarrow WW$
for (a) the $\mtjj$ analysis and (b) the $\etjj$ analysis. 
The solid (dashed) lines show the sum of all production processes with 
$H\rightarrow \tau\tau$ ($H\rightarrow WW$) decays at $M_H=125$ GeV.
Events near BDT$_{\rm TW}=-1$ are dominantly $H\rightarrow WW$
and those near $+1$ are mainly $H\rightarrow\tau\tau$.  
}
\end{center}
\end{figure}

We perform a separation of the data, SM MC backgrounds and MJ background into
two subsamples: one constructed to be rich in $H\rightarrow\tau\tau$
signals (the T subsample), and another rich in $H\rightarrow WW$ signals 
(the W subsample).  We perform this 
separation into orthogonal data sets with a BDT (BDT$_{\rm TW}$) based on the variables listed in 
Table~\ref{tab-bdtinputs} using a TMVA training between the $H\rightarrow \tau\tau$ and 
$H\rightarrow WW$  MC signal
events.  Each subsample subsequently undergoes its own multivariate analysis and the
results are combined at the limit-setting stage.  The subsample separation 
gives about 15 -- 20\% improvement in final Higgs boson limits over the no-separation case.
It also provides an analysis for a purified $H\rightarrow\tau\tau$ signal, 
thus giving additional information on fermionic decays of the Higgs boson.
We employ the BDT$_{\rm TW}$ only to define the
T and W subsamples; no further use of this variable is made.
The BDT$_{\rm TW}$ distributions for $\mtjj$ and $\etjj$ are shown in 
Fig.~\ref{fig-bdttauw}. 
  
We impose a cut to separate the T and W subsamples
at BDT$_{\rm TW}=+0.3$ for the $\mtjj$ analysis 
and at $-0.6$ for the $\etjj$ analysis.
The purity of $\tau\tau$ decays in the T subsample and of
$WW$ decays in the W subsample are about 90\% in the regions
where the respective signals dominate.
The dominant backgrounds in the T subsample are $\zj$ and (for the $\mtjj$ analysis only) MJ. 
For the W subsample, the dominant backgrounds are MJ, $t\overline t$, $\wj$ and 
(for the $\etjj$ analysis only) $Z(ee)+$jets.


\subsection{\label{sec-globalbdt}Global BDTs}

In searches for the Higgs boson, separate multivariate analyses
have generally been performed for each Higgs boson mass under consideration.
Due to variations in the details in the BDT training,
this can lead to fluctuations in limits from one mass point to another.
We have constructed a method that reduces such unwanted fluctuations.

We first perform a {\it single} training using the Higgs boson
signal MCs at {\it all} mass points (``global BDT").  
In this training, each signal event is characterized by 
the set of variables given in Table~\ref{tab-bdtinputs}, including the
value of $M_H$ appropriate to each specific signal sample.  To prevent the
classifier from artificially separating backgrounds from signals based on the 
$M_H$ value, we randomly assign a $M_H$ value to the backgrounds with
a $M_H$ distribution constructed to reproduce that of the signal samples.
This assignment is done separately for the T and W subsamples 
and takes into account that there is an admixture of $H\rightarrow \tau\tau$
and $H\rightarrow WW$ decays in both cases. The $M_H$ variable does not play a strong
role in separating signals from backgrounds in the global BDT training.  

The global BDT is characterized by a 
set of training weights for the splitting of signal and backgrounds at
each of the nodes of the tree.  We then form the final discriminant at each $M_H$
by passing the MC signal events for a particular $M_H$ through the
weights provided by the global BDT, now with $M_H$ set to the value under
consideration.  In this pass, the backgrounds are also provided with this 
specific $M_H$ value, and the data events to which the distributions are to
be compared are similarly provided with the test $M_H$ value.  
The global BDT approach removes the variation in the training at different $M_H$
values and provides a more uniform distribution of Higgs boson cross section limits
with minimal ($\approx 10\%$) deterioration in limits relative to 
separate training at each mass point.

\subsection {\label{sec-bin}Choice of BDT binning}

If the subsample BDT distributions have bins with a small number
of background events, there can be statistical fluctuations in the calculated limits.
These are reduced by choosing BDT bin sizes that ensure that all bins have at
least 20 background events before application of event weights.  

\begin{figure}[t]
\begin{center}
\includegraphics[width=0.400\textwidth]{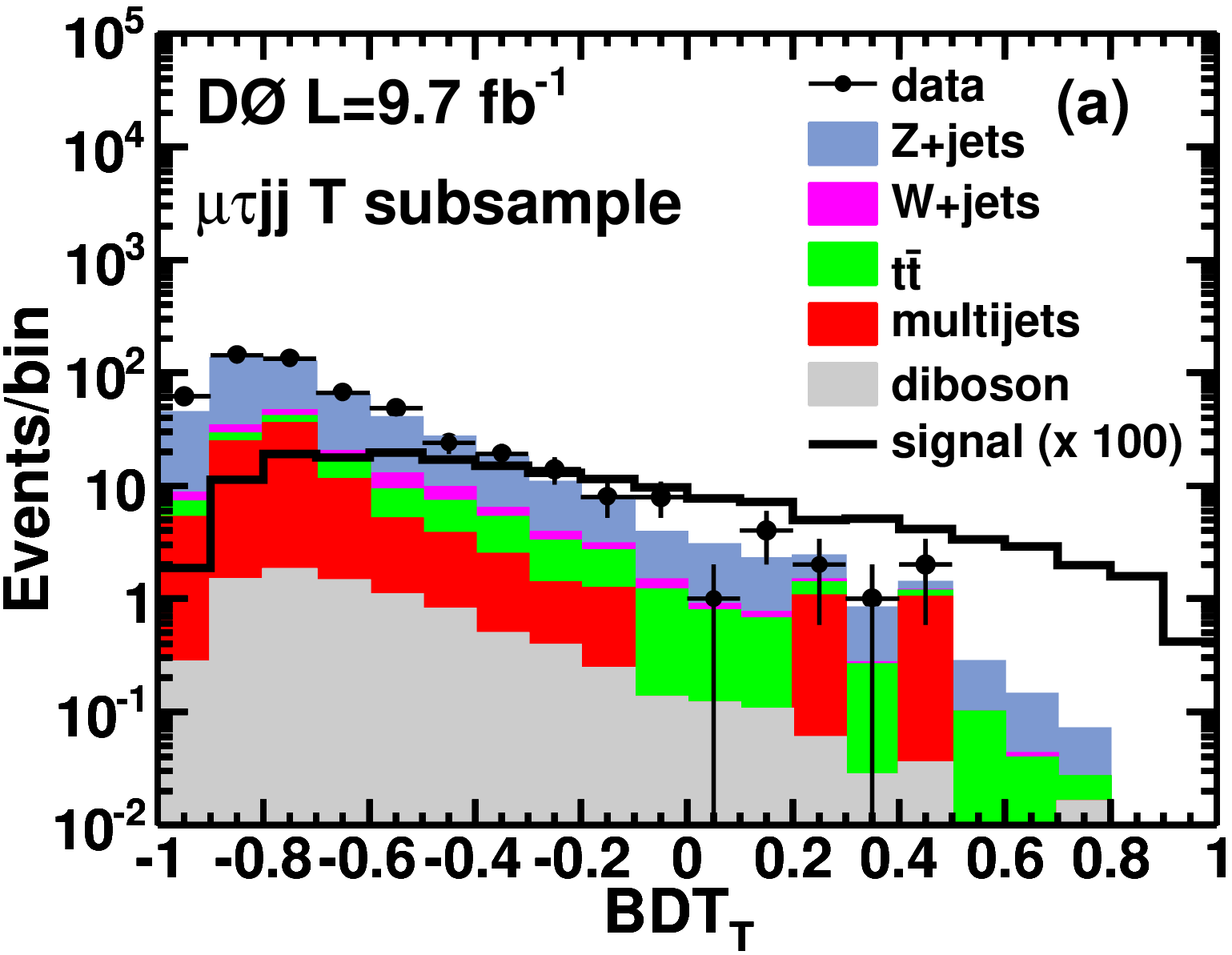} \\
\includegraphics[width=0.400\textwidth]{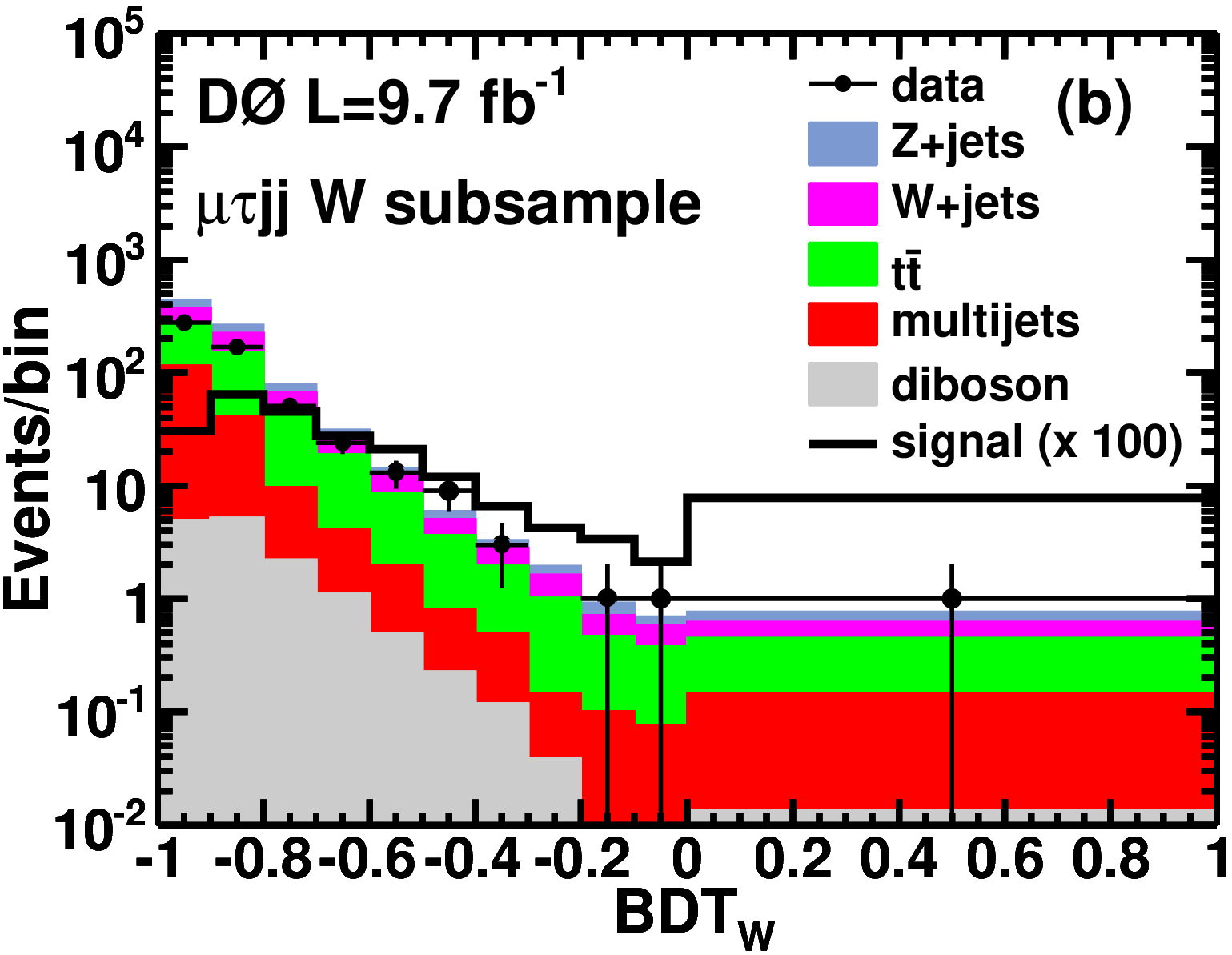} \\
\caption{\label{figm_finalbdt}
(color online) 
Final discriminant distributions for the $\mtjj$ analysis
for (a) the T subsample at $M_H=125$ GeV and 
(b) the W subsample at $M_H=145$ GeV.
}
\end{center}
\end{figure}

\begin{figure}[t]
\begin{center}
\includegraphics[width=0.400\textwidth]{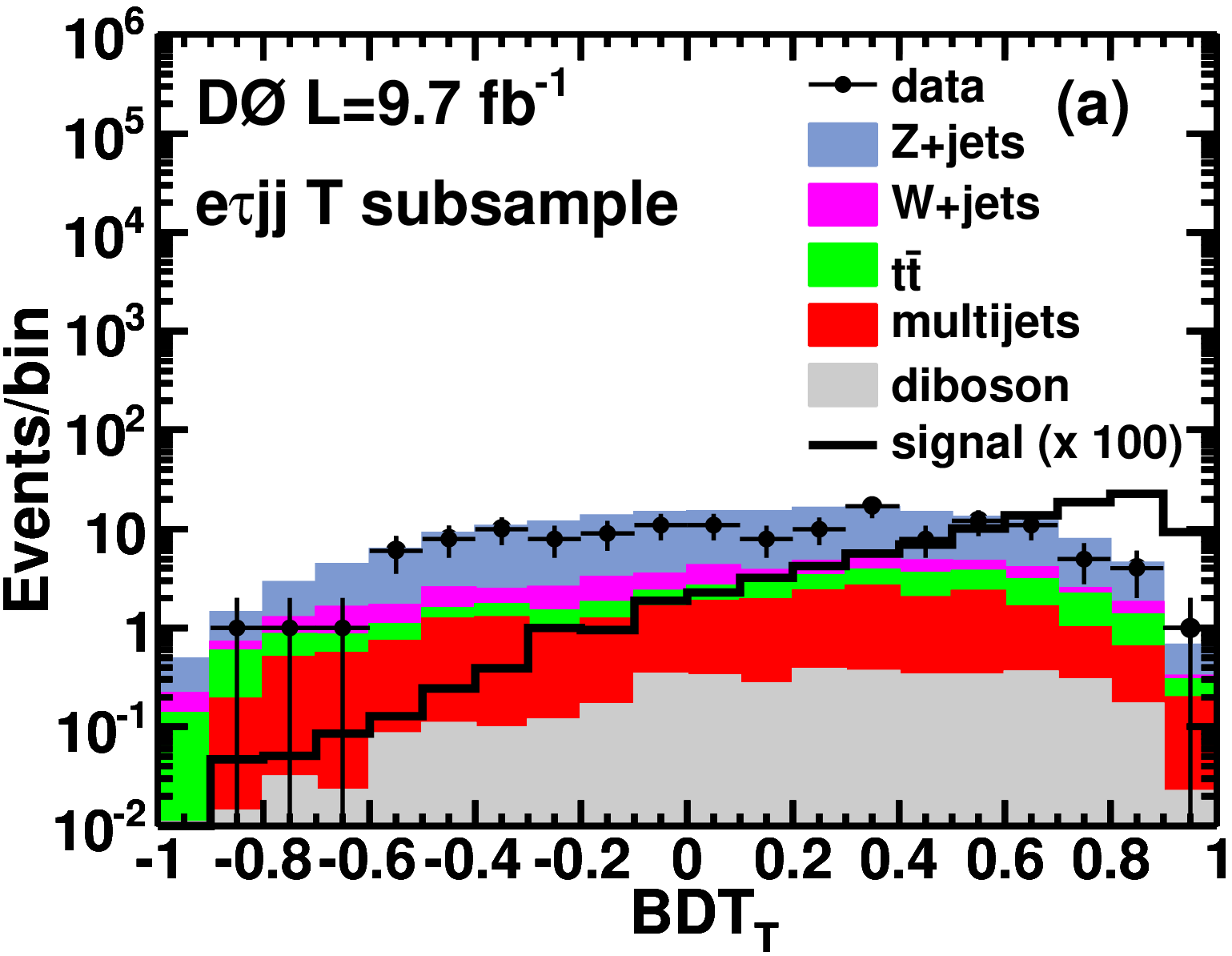} \\
\includegraphics[width=0.400\textwidth]{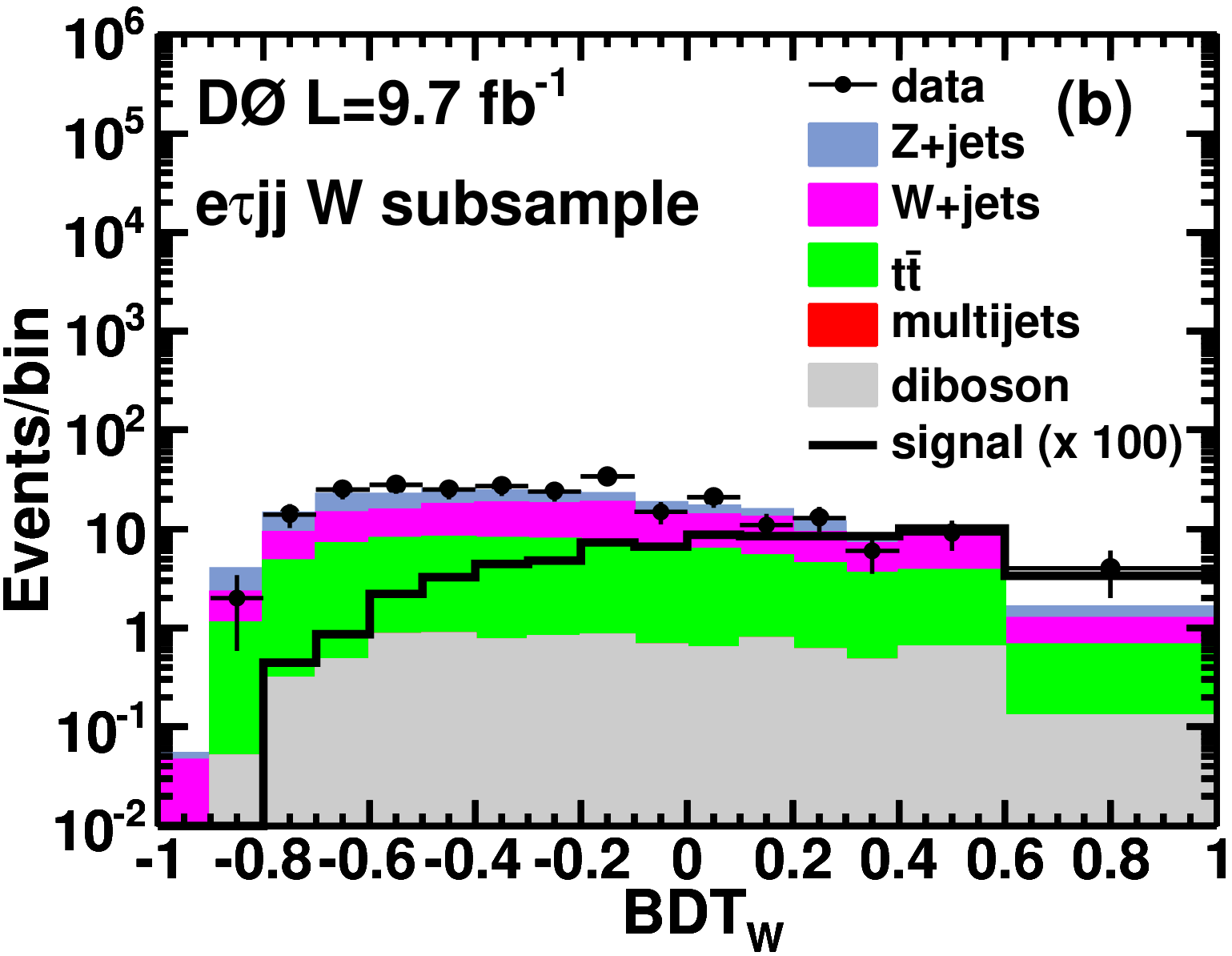} \\
\caption{\label{fige_finalbdt}
(color online) 
Final discriminant distributions for the $\etjj$ analysis
for (a) the T subsample at $M_H=125$ GeV and 
(b) the W subsample at $M_H=145$ GeV.
}
\end{center}
\end{figure}

Representative final discriminants are shown for the T subsample (at $M_H=125$ GeV)
and the W sample (at $M_H=145$ GeV) in Figs.~\ref{figm_finalbdt} and \ref{fige_finalbdt}
for the $\mtjj$ and $\etjj$ analyses respectively.  The agreement of predicted backgrounds 
and the observed data is good.   


\section {\label{sec-syst}Systematic Uncertainties}

A large number of systematic uncertainties have been considered, often
broken down separately by analysis channel and subsample, 
tau type, or background or signal process.  
The luminosity and trigger uncertainties are obtained from
separate analyses of D0 data.  
The lepton and tau identification uncertainties are obtained
from special samples enriched in $Z$ boson decays.  
The jet energy scale, energy resolution, 
identification and vertex confirmation uncertainties are obtained from
special dijet and $\gamma$+jet samples separately for the T and W subsamples.
Uncertainties
in the SM background cross section normalizations and shapes are obtained using theoretical
uncertainties, and the extent to which special data samples enriched in each background process
agree with MC predictions.  The MJ background uncertainties are determined
by comparing the results using the MJ-enriched samples with
those obtained using the SS signal sample after SM background subtraction.   
Signal cross section uncertainties are obtained
from theoretical estimates and include the effect of PDF uncertainties.
Table~\ref{tab-syst} summarizes the systematic uncertainties on the various sources.
For each source, the impact on the final discriminant 
is assessed by changing the appropriate parameters by 1 s.d from the nominal values.
Some of the uncertainties affect only the normalization of the final discriminant distribution
and some modify its shape while keeping the normalization fixed.


\begin{table}[htbp]
\caption{\label{tab-syst}
The range of systematic uncertainties (in percent) from different sources. 
Type N (S) denote normalization (shape) uncertainties on the final 
discriminant.
``XS'' denotes ``cross section''. 
The ``jets'' uncertainties represent the independent uncertainties
arising from jet vertex confirmation,
jet identification and efficiency, jet energy resolution and jet energy
scale which, for a given channel and subsample, are similar.
}
\begin{tabular}{lcc} \hline \hline
Source & Type & Uncertainty (\%) \\ \hline
Luminosity                       & N & 6.1   \\
$\mu$ ID/track match/isolation   & N & 2.9   \\
$e$ ID/isolation                 & N & 4.0   \\
Single $\mu$ trigger efficiency  & N & 5     \\
All trigger/Single $\mu$ trigger & N & 7     \\
$e+$Jets trigger efficiency      & N & 2     \\
$\tau_h$ selection (by type)       & N & 5.5/4.0/6.0 \\
$\tau_h$ energy scale              & N & 9.8   \\
$\tau_h$ track efficiency          & N & 1.4   \\
$W/Z$+jets XS                    & N & 6.0   \\
$t\overline t$, single top XS    & N & 7.0   \\
diboson XS                       & N & 6.0   \\
VH signal XS                     & N & 6.2   \\
VBF signal XS                    & N & 4.9   \\
GF signal XS normalization      & N & 33    \\
GF signal XS PDF                & N & 29    \\
jets $\mtjj$ T (W) subsample     & S & 2 -- 11 (1 -- 11) \\
jets $\etjj$ T (W) subsample     & S & 4 -- 20 (2 -- 15) \\
PDF (signals)                    & N & 1.6   \\
PDF (backgrounds)                & N & 2.0   \\
$\mtjj$ MJ normalization         & N & 5.3   \\
$\etjj$ MJ normalization         & N & 5.0   \\
$\mtjj$ MJ shape                 & S & 5 -- 10 \\  \hline \hline
\end{tabular}
\end{table}

\section{\label{sec-limits}Cross Section Limits}

\begin{figure*}[t]
\begin{center}
\includegraphics[width=0.320\textwidth]{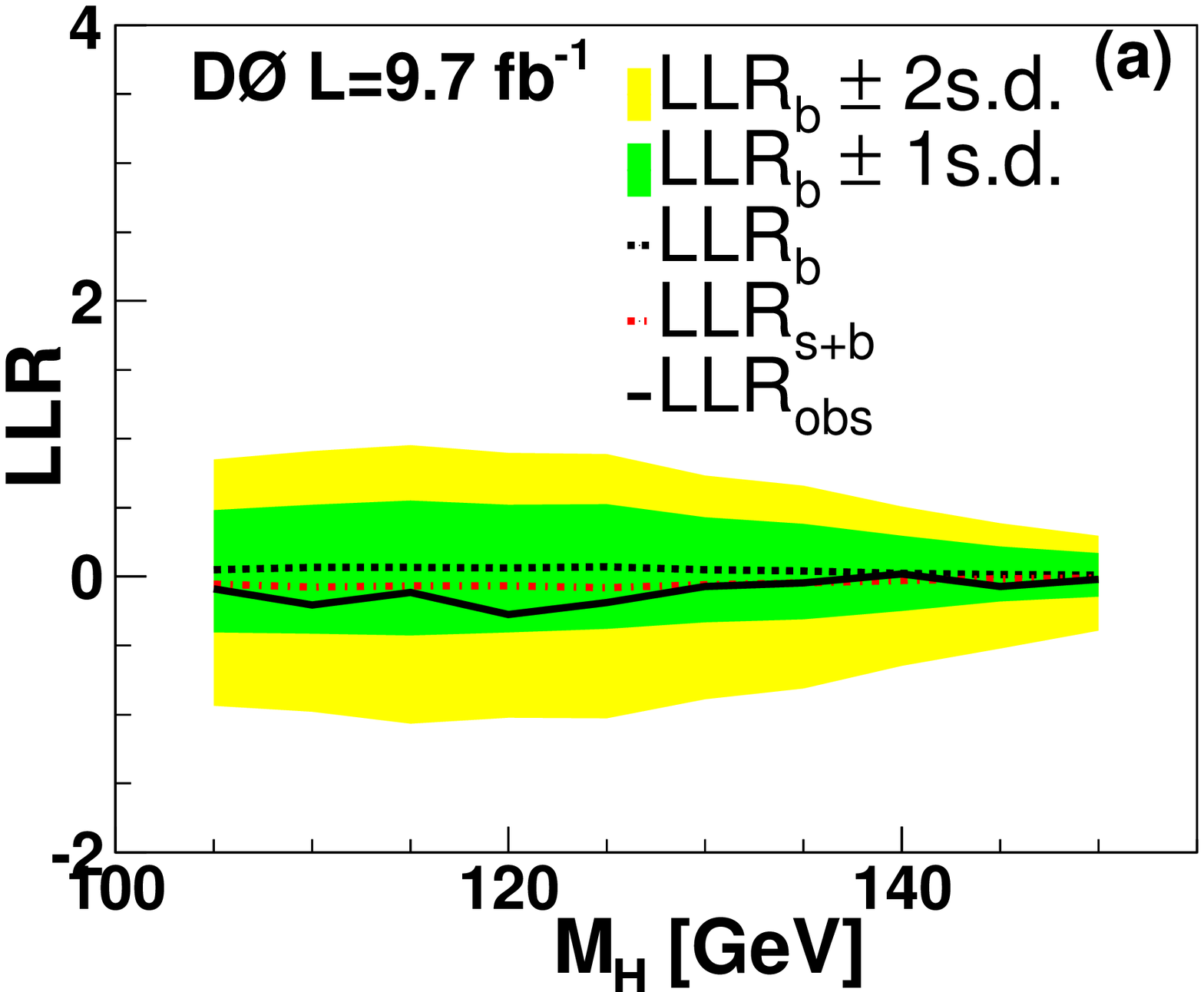}
\includegraphics[width=0.320\textwidth]{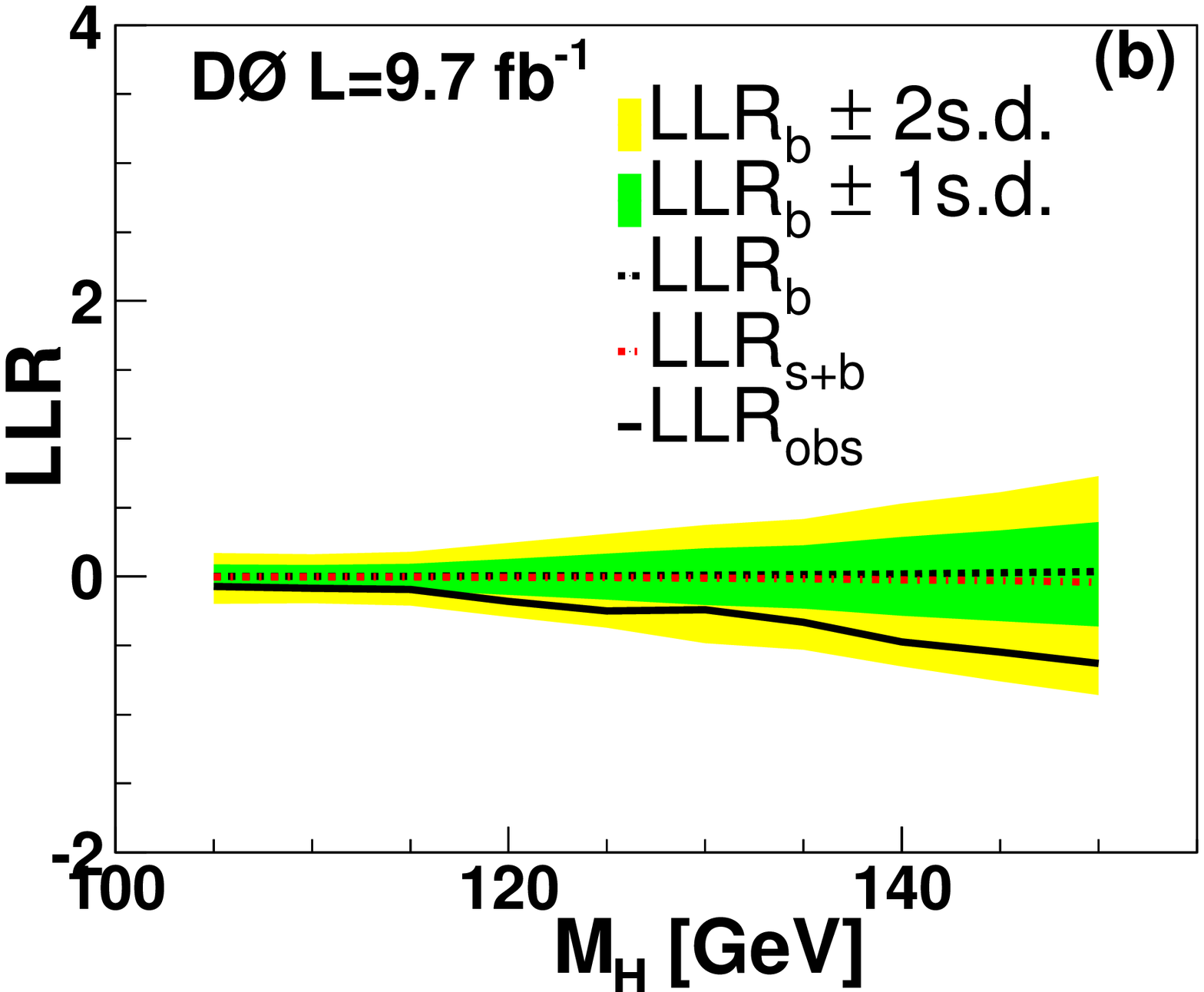} 
\includegraphics[width=0.320\textwidth]{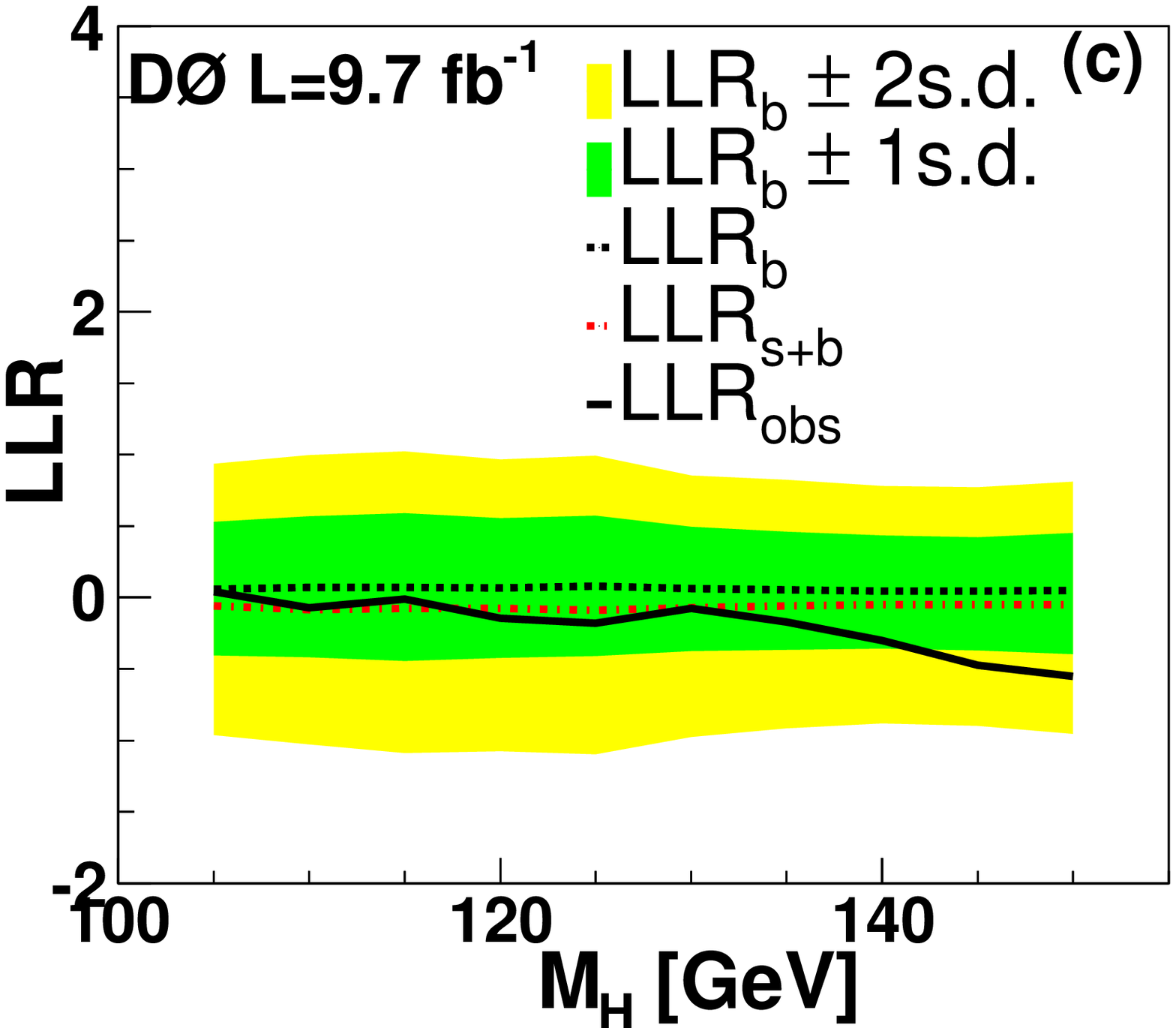} \\
\caption{\label{fig-LLR}
(color online) 
LLR distributions for (a) the T subsample, (b) the W subsample, and (c)
the sum of the combined T and W subsamples.   The $\mtjj$ and $\etjj$ analyses are
combined.  The black dashed line shows the expected LLR for the 
background-only hypothesis and the red dash-dotted line shows the
expected LLR for the signal-plus-background hypothesis.  The solid black line 
indicates the observed LLR.  The green (yellow) shaded bands indicate the 
$\pm$1 s.d. ($\pm$2 s.d.) uncertainties on the expected background only LLR.
}
\end{center}
\end{figure*}

The upper limits on the Higgs boson cross section for each analysis are obtained
from the final discriminants for Higgs boson masses between 105 and 150 GeV in 5 GeV increments
obtained with the modified frequentist method of Ref.~\cite{modfreq}, using a 
negative log likelihood ratio (LLR) for
the background-only and signal-plus-background hypotheses as the test statistic.   
The LLR plots, individually for T and W subsamples
and for their sum, are shown in Fig.~\ref{fig-LLR}.


\begin{figure*}[t]
\begin{center}
\includegraphics[width=0.320\textwidth]{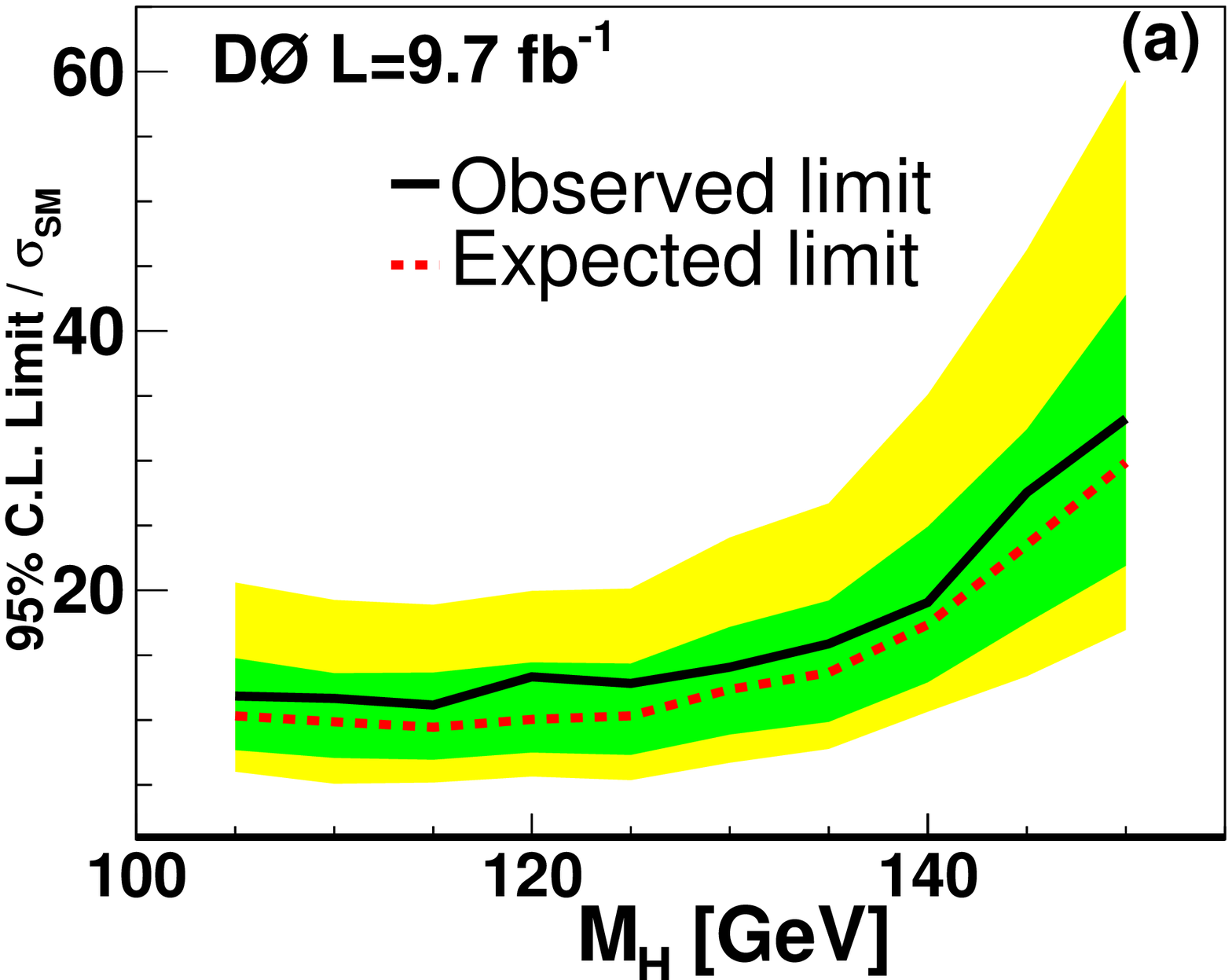}
\includegraphics[width=0.320\textwidth]{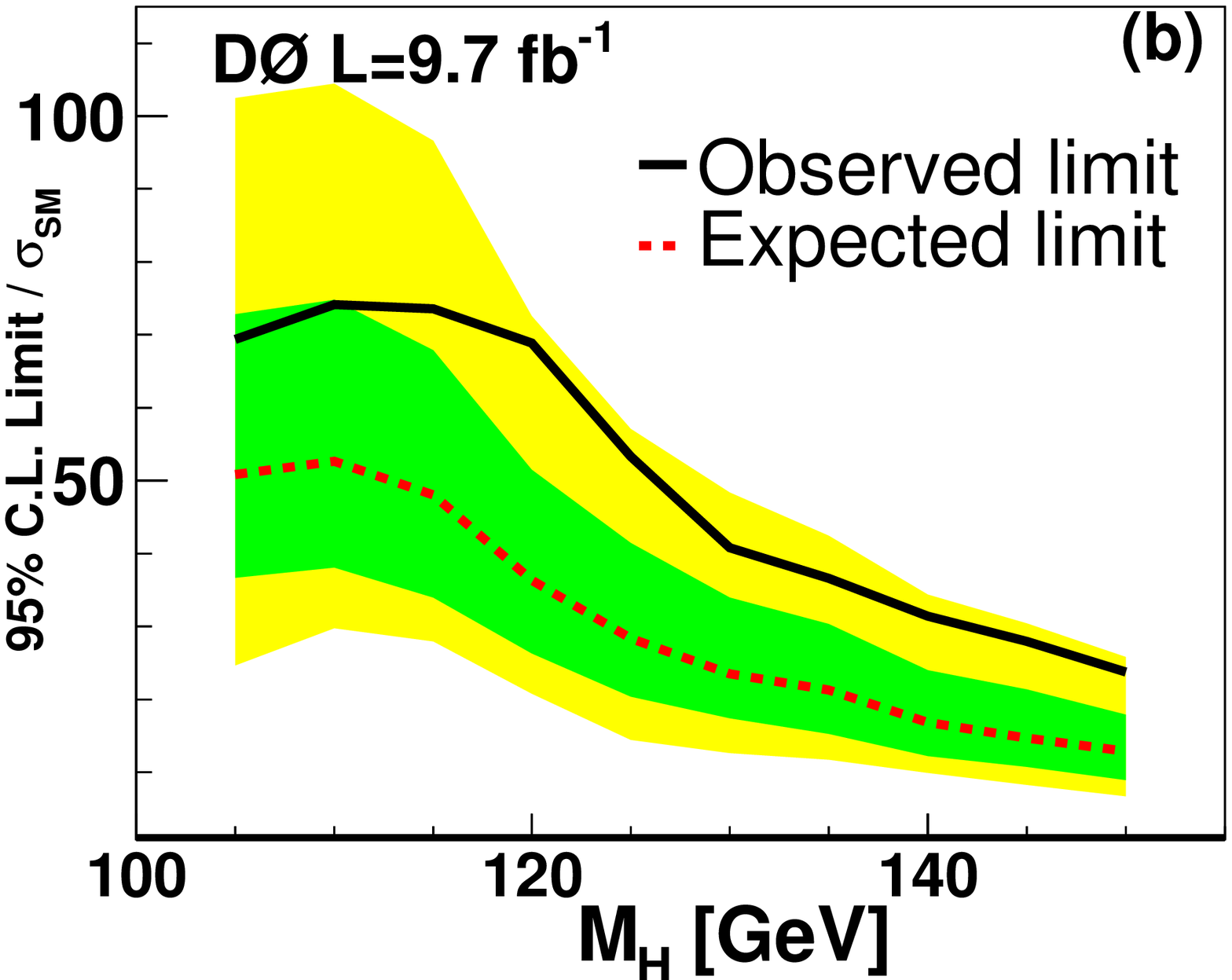}
\includegraphics[width=0.320\textwidth]{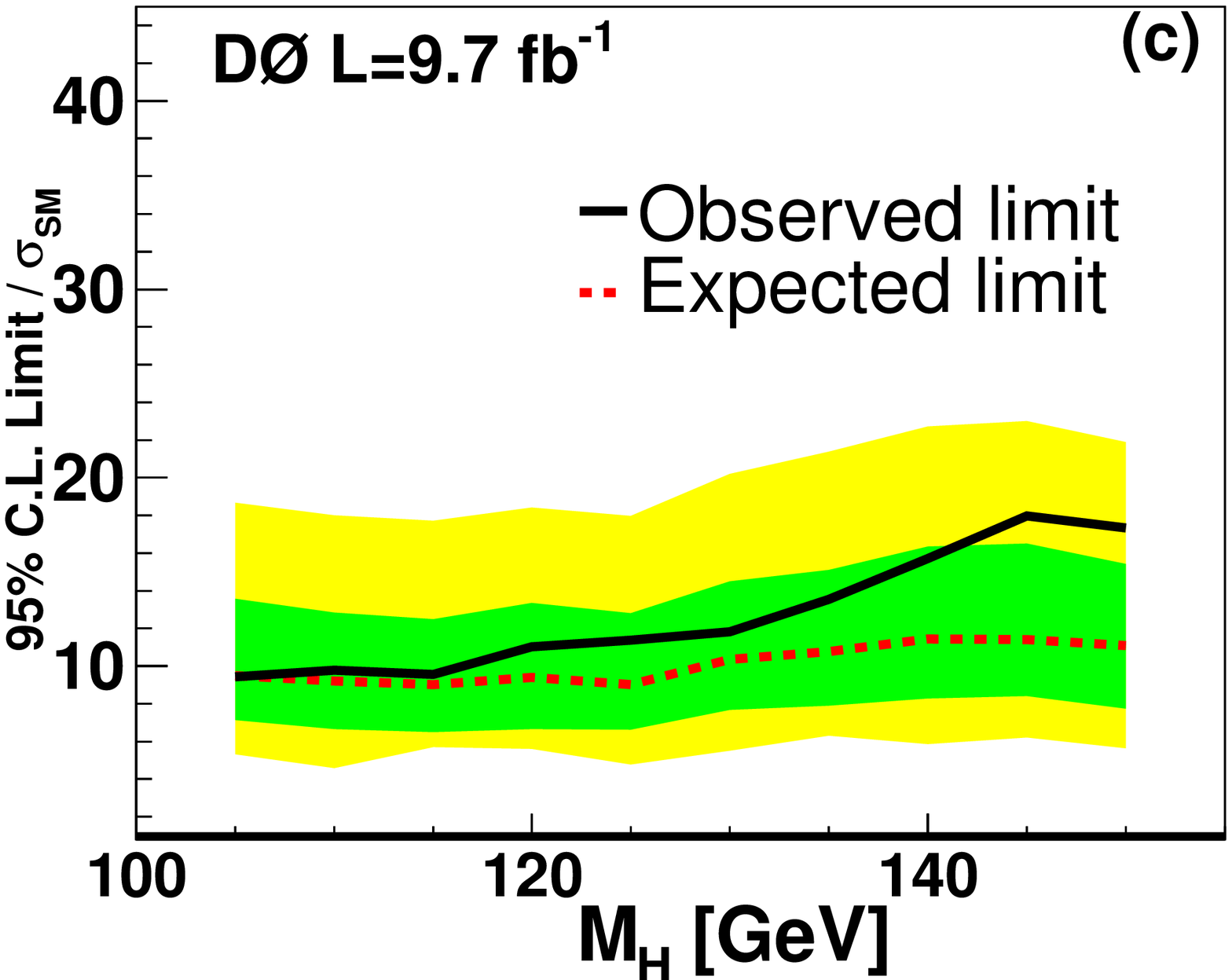} \\
\caption{\label{fig-limits}
(color online) 
The ratio of 95\% C.L. upper limits on Higgs boson production for the
(a) T subsample, (b) W subsample, and 
(c) the sum of the T and W subsamples.
The $\mtjj$ and $\etjj$ analyses are combined. 
The green (yellow) shaded bands indicate the 
$\pm$1 s.d. ($\pm$2 s.d.) uncertainties on the expected limit ratios.
}
\end{center}
\end{figure*}

The impact of systematic uncertainties on the limits is reduced by maximizing a ``profile''
likelihood function~\cite{wade} in which these uncertainties are constrained to Gaussian priors
to give a best fit to the data.
The appropriate correlations are retained (for example,
the VH cross section uncertainty is fully correlated across the $\mtjj$ and $\etjj$ analyses
and the T and W subsamples).
The value of the Higgs boson cross section is adjusted in each limit calculation until the value
of $CL_s$ reaches 0.05, corresponding to the 95\% C.L., where $CL_s = CL_{s+b}/CL_b$ and $CL_{s+b}$
($CL_b$) are the probabilities for the negative LLR value observed in 
simulated signal+background (background) pseudo-experiments to be less signal-like than that observed
in our data.  The ratio of the resulting 95\% C.L. upper limits 
to the SM predictions on production times branching ratio are shown in 
Fig.~\ref{fig-limits}, and in Table~\ref{tab-limits}.  


\begin{table*}[htbp]
\caption{\label{tab-limits}
The ratio of 95\% C.L. upper limits on Higgs boson production to the
SM prediction.
}

\begin{tabular}{crrrrrr|rrrrrr|rrrrrr} \hline \hline
~~~~$M_H$~~ & 
\multicolumn{6}{c|}{$\mtjj$} & 
\multicolumn{6}{c|}{$\etjj$} &
\multicolumn{6}{c}{$\mtjj~+~\etjj$}  \\ \hline 
 & \multicolumn{2}{c}{T} & \multicolumn{2}{c}{W} & \multicolumn{2}{c|}{T+W} 
 & \multicolumn{2}{c}{T} & \multicolumn{2}{c}{W} & \multicolumn{2}{c|}{T+W} 
 & \multicolumn{2}{c}{T} & \multicolumn{2}{c}{W} & \multicolumn{2}{c}{T+W}  \\ \hline
 & \multicolumn{1}{c}{exp} &  \multicolumn{1}{c}{obs}
 & \multicolumn{1}{c}{exp} &  \multicolumn{1}{c}{obs}
 & \multicolumn{1}{c}{exp} &  \multicolumn{1}{c|}{obs}
 & \multicolumn{1}{c}{exp} &  \multicolumn{1}{c}{obs}
 & \multicolumn{1}{c}{exp} &  \multicolumn{1}{c}{obs}
 & \multicolumn{1}{c}{exp} &  \multicolumn{1}{c|}{obs}
 & \multicolumn{1}{c}{exp} &  \multicolumn{1}{c}{obs}
 & \multicolumn{1}{c}{exp} &  \multicolumn{1}{c}{obs}
 & \multicolumn{1}{c}{exp} &  \multicolumn{1}{c}{obs} \\ \hline
105 & 15.3 & 18.4 & 116  & 84.9 & 14.1 & 16.3~ & ~16.3 & 14.6 & 62.3 & 107  & 14.9 & 11.8~ & ~10.3 & 11.8 & 50.9 & 69.3 & 9.4  & 9.4  \\
110 & 14.3 & 18.2 & 113  & 112  & 13.9 & 16.8~ & ~15.6 & 14.6 & 64.3 & 101  & 14.4 & 12.5~ & ~9.9  & 11.7 & 52.6 & 74.1 & 9.1  & 9.8  \\
115 & 14.6 & 17.3 & 86.6 & 121  & 13.8 & 17.2~ & ~14.9 & 14.8 & 66.5 & 93.3 & 14.2 & 11.7~ & ~9.5  & 11.2 & 48.0 & 73.5 & 9.0  & 9.5  \\
120 & 15.5 & 16.5 & 60.1 & 125  & 14.6 & 17.6~ & ~15.7 & 19.7 & 56.6 & 75.4 & 14.5 & 15.3~ & ~10.1 & 13.3 & 36.4 & 68.9 & 9.4  & 11.1 \\
125 & 15.3 & 17.0 & 48.9 & 84.9 & 14.4 & 18.4~ & ~15.9 & 18.7 & 46.1 & 65.0 & 14.6 & 16.0~ & ~10.4 & 12.8 & 28.4 & 53.3 & 9.0  & 11.3 \\
130 & 17.1 & 18.3 & 32.8 & 55.4 & 15.3 & 17.4~ & ~20.1 & 21.8 & 51.5 & 65.3 & 18.3 & 19.5~ & ~12.4 & 14.1 & 23.5 & 40.8 & 10.2 & 11.8 \\
135 & 19.1 & 19.9 & 28.8 & 53.7 & 15.6 & 21.0~ & ~23.6 & 25.6 & 51.0 & 51.8 & 20.4 & 21.4~ & ~13.7 & 15.9 & 21.3 & 36.6 & 10.8 & 13.5 \\
140 & 24.3 & 23.7 & 22.5 & 44.5 & 15.9 & 22.8~ & ~31.0 & 33.9 & 40.0 & 40.8 & 24.4 & 26.3~ & ~17.4 & 19.1 & 16.9 & 31.4 & 11.5 & 16.0 \\
145 & 30.7 & 33.9 & 19.1 & 37.3 & 15.1 & 23.7~ & ~42.5 & 47.6 & 45.5 & 44.0 & 29.8 & 36.0~ & ~23.5 & 27.6 & 14.7 & 27.9 & 11.5 & 17.9 \\
150 & 38.1 & 36.0 & 17.5 & 34.8 & 14.9 & 22.2~ & ~58.4 & 69.5 & 29.0 & 27.5 & 24.8 & 30.8~ & ~29.9 & 33.3 & 12.9 & 23.7 & 11.1 & 17.2 \\ \hline \hline
\end{tabular}
\end{table*}

In summary we have searched for the SM Higgs boson in final states involving an electron or muon
and a hadronically decaying tau, together with at least two jets.  We set 
95\% C.L. limits on the ratio of
the Higgs boson production cross section to that predicted in the SM of 11.3 times 
for a Higgs boson mass of 125 GeV, to be compared with an expected ratio of 9.0.
For a subsample enriched in $H\rightarrow \tau\tau$ decays we observe (expect) a ratio
of 12.8 (10.4) for $M_H=125$ GeV.  These are the most stringent limits
on Higgs boson production with $H\rightarrow \tau\tau$ decay at the Tevatron to date.
For an orthogonal subsample enriched in $H\rightarrow WW$ decays
the corresponding ratios are 14.7 (11.5) at $M_H=145$ GeV.

\vspace*{2.5mm}

%
We thank the staffs at Fermilab and collaborating institutions,
and acknowledge support from the
DOE and NSF (USA);
CEA and CNRS/IN2P3 (France);
MON, Rosatom and RFBR (Russia);
CNPq, FAPERJ, FAPESP and FUNDUNESP (Brazil);
DAE and DST (India);
Colciencias (Colombia);
CONACyT (Mexico);
NRF (Korea);
FOM (The Netherlands);
STFC and the Royal Society (United Kingdom);
MSMT and GACR (Czech Republic);
BMBF and DFG (Germany);
SFI (Ireland);
The Swedish Research Council (Sweden);
and
CAS and CNSF (China).

\end{document}